\documentclass[preprint,review,12pt]{elsarticle}

\usepackage{amsmath, amssymb, amscd, amsthm, amsfonts, graphicx, hyperref, natbib,array,xcolor}
\newcolumntype{P}[1]{>{\centering\arraybackslash}p{#1}}
\newcolumntype{M}[1]{>{\centering\arraybackslash}m{#1}}
\usepackage[margin=2.3cm]{geometry}

\journal{Mathematical Biosciences}

\begin{document}

\begin{frontmatter}



\title{A new lipid-structured model to investigate the opposing effects of LDL and HDL on atherosclerotic plaque macrophages}


\author[Oxford]{Keith L Chambers\corref{cor}}
\ead{keith.chambers@maths.ox.ac.uk}
\affiliation[Oxford]{organization={Wolfson Centre for Mathematical Biology, Mathematical Institute, University of Oxford},%Department and Organization
            addressline={Andrew Wiles Building, Radcliffe Observatory Quarter, Woodstock Road}, 
            city={Oxford},
            postcode={OX2 6GG}, 
            state={Oxfordshire},
            country={United Kingdom}
            }
\cortext[cor]{Corresponding author}

\author[Sydney]{Mary R Myerscough}
\ead{mary.myerscough@sydney.edu.au}

\author[Oxford]{Helen M Byrne}
\ead{helen.byrne@maths.ox.ac.uk}

\affiliation[Sydney]{organization={School of Mathematics and Statistics, University of Sydney},%Department and Organization
            addressline={Carslaw Building, Eastern Avenue, Camperdown}, 
            city={Sydney},
            postcode={2006}, 
            state={New South Wales},
            country={Australia}}

\begin{abstract}

Atherosclerotic plaques %are a major cause of heart attacks and strokes. They 
form in artery walls due to a chronic inflammatory response driven by lipid accumulation. A key component of the inflammatory response is the interaction between monocyte-derived macrophages and extracellular lipid. Although concentrations of low-density lipoprotein (LDL) and high-density lipoprotein (HDL) particles in the blood are known to affect plaque progression, their impact on the lipid load of plaque macrophages remains unexplored. In this paper, we develop a lipid-structured mathematical model to investigate the impact of blood LDL/HDL levels on plaque composition, and lipid distribution in plaque macrophages. A reduced subsystem, derived by summing the equations of the full model, describes the dynamics of biophysical quantities relating to plaque composition (e.g. total number of macrophages, total amount of intracellular lipid). We also derive a continuum approximation of the model to facilitate analysis of the macrophage lipid distribution. The results, which include time-dependent numerical solutions and asymptotic analysis of the unique steady state solution, indicate that plaque lipid content is %particularly 
sensitive to the influx of LDL relative to HDL capacity. The macrophage lipid distribution evolves in a wave-like manner towards %one of several qualitatively distinct steady state solutions. The 
an equilibrium profile which may be monotone decreasing, %(which may be concave or convex)
quasi-uniform or unimodal, attaining its maximum value at a non-zero lipid level. Our model also reveals that
%, since macrophages have a finite capacity, 
macrophage uptake may be severely impaired by lipid accumulation. We conclude that lipid accumulation in plaque macrophages may serve as a partial explanation for the defective uptake of apoptotic cells (efferocytosis) often reported in atherosclerotic plaques.

\end{abstract}


% \begin{highlights}
% \end{highlights}

\begin{keyword}
 Atherosclerosis \sep Lipid  \sep Macrophage \sep LDL \sep HDL \sep Structured population model
% %% MSC codes here, in the form: \MSC code \sep code
% %% or \MSC[2008] code \sep code (2000 is the default)
% \MSC 9210 \sep 1111
\end{keyword}

\end{frontmatter}


\section{Introduction} \label{sec: intro}

Atherosclerosis is a chronic inflammatory disease that develops in the walls of major arteries \cite{back2019inflammation,tabas2016macrophage}. The condition begins when disturbed blood flow allows fatty compounds called lipids, which are carried upon low-density lipoprotein (LDL) particles, to enter the wall from the bloodstream \cite{boren2020low}. The accumulation of lipid in the wall triggers an immune response that attracts monocyte-derived macrophages (MDMs) to the lesion. Macrophages are the dominant immune cell by number in atherosclerotic lesions \cite{razeghian2022immune}. They play a key role in disease progression by ingesting extracellular lipid in a process called \textit{phagocytosis}. Over time the lesion may develop into an atherosclerotic plaque with a large core of extracellular lipid due to the persistent influx of lipid and the death of lipid-laden macrophages \cite{guyton1996development}. The rupture of such a plaque releases the lipid core material into the bloodstream, where it promotes blood clot formation. Blood clots can, in turn, block vessels in the heart and brain and induce myocardial infarction or stroke. The mechanisms that underpin macrophage lipid accumulation and extracellular lipid core formation remain an active area of research.

% Blood lipoprotein concentrations are known to significantly impact the development of atherosclerosis \cite{kohsaka2010relationship, badimon2012ldl, sun2022predictive, hao2014ldl}. Lipoprotein particles can be divided into two main types which have opposing effects on disease progression: low-density lipoprotein (LDL, `bad cholesterol') and high-density lipoprotein (HDL `good cholesterol'). LDL particles drive atherosclerosis as a major source of lesion lipid and consequently MDMs via recruitment. Recruited macrophages clear the extracellular space of lipid via phagocytosis but 

% MDM recruitment provides a secondary source of lesion lipid since macrophages contain 

Macrophage lipid content is the product of phagocytosis and offloading events. Macrophages increase their lipid load by clearing the extracellular environment of LDL particles (which become modified upon entry and susceptible to macrophage ingestion \cite{alique2015ldl}) and cellular debris via phagocytosis. In particular, macrophages phagocytose the lipid-rich contents of nearby dead macrophages. This includes lipid ingested by dead cells during their lifetime and also endogenous lipid contained in the membranes of cell corpses. Macrophages reduce their lipid burden by offloading lipid onto high-density-lipoprotein (HDL) particles, which also enter the lesion from the bloodstream. The lipoproteins LDL and HDL (which carry `bad' and `good' cholesterol respectively) therefore have opposing effects on macrophage lipid content. The impact of blood LDL/HDL levels on overall plaque composition and the distribution of lipid within the MDM population are well known but not completely understood.

The efficiency of macrophage phagocytosis and offloading is modulated by intracellular signalling  \cite{zent2017maxed, remmerie2018macrophages}). We note firstly that macrophages have a finite capacity for phagocytosis \cite{zent2017maxed}. More specifically, experimental observations indicate that macrophages surrounded by an abundance of phagocytic targets dynamically regulate their rate of uptake according to the available capacity. This behaviour is the result of physical limitations (e.g. finite cell size) and intracellular biochemical negative feedback signals. The study of Pinney et al. \cite{pinney2020macrophage} indicates that the down-regulation of phagocytosis occurs on a rapid timescale  ($< 1$ hour). Hence, the reduction in macrophage phagocytic efficiency is likely to be commensurate with the current phagocytic load of a cell. Similarly, the efficiency of lipid offloading is also mediated by intracellular signalling \cite{remmerie2018macrophages}. The accumulation of cholesterol is known to activate transcription factors (e.g. Lxr$\alpha$, Lxr $\beta$) that upregulate the expression of transporters (e.g. ABCA1) which mediate cholesterol efflux to HDL particles. Taken together, the observations above indicate that, on average, heavily lipid-laden macrophages are less efficient at phagocytosis but more efficient at offloading than macrophages with a lesser lipid burden. Interestingly, many studies report that macrophage phagocytosis of apoptotic cells (termed \textit{efferocytosis}) is defective in mid-to-late atherosclerotic plaques (\cite{kojima2017role,yurdagul2018mechanisms}). However, the extent to which this defective efferocytosis is attributable to the intrinsic down-regulation of phagocytosis due to lipid loading is not known.

%Needs updating!
There is growing interest in mathematical models of atherosclerosis \cite{avgerinos2019mathematical,el2019mathematical}. Many models partition the plaque macrophages into two sub-populations: those with a low lipid load (termed `macrophages') and those that are heavily lipid-laden (termed `foam cells') \cite{calvez2009mathematical,chalmers2017nonlinear,silva2020modeling}. Lipid-dependent rates of phagocytosis and offloading are incorporated in this framework by assuming that only macrophages can ingest lipid and only foam cells can offload lipid (e.g. \cite{chalmers2017nonlinear}). A more natural approach of representing lipid accumulation in plaque macrophages is to use a structured population model. Lipid-structured models for atherosclerosis have been developed by several authors \cite{ford2019lipid,chambers2022lipid,watson2022lipid,meunier2019mathematical}. The models of Ford et al. \cite{ford2019lipid}, Chambers et al. \cite{chambers2022lipid} and Watson et al. \cite{watson2022lipid} assume, for simplicity, that the rates of lipid uptake and offloading are independent of lipid load. The model of Meunier and Muller \cite{meunier2019mathematical} accounts for an arbitrary lipid-dependent uptake rate of LDL, but does not account for uptake of cellular debris nor lipid offloading. 

In this paper we present a new lipid-structured model for macrophage populations in atherosclerotic plaques. The primary aims of the study are to:
\begin{enumerate}
    \item Investigate how blood LDL/HDL concentrations impact 
    \subitem (a) Plaque composition, and 
    \subitem (b) The distribution of lipid in plaque macrophages;
    \item Investigate the extent to which defective phagocytosis can be attributed to high lipid loads in macrophages. 
\end{enumerate} 
The model is novel in that it (i) accounts for lipid-dependent rates of phagocytosis and offloading, (ii) explicitly tracks the amounts of extracellular lipid and HDL lipid capacity in the plaque, and (iii) is fundamentally based on a discrete structuring of the macrophage population. We also derive a continuum approximation of the discrete model to facilitate analysis of the macrophage lipid distribution in Section \ref{sec: lipid distribution}. The continuum approximation replaces the dynamical equations of the macrophage population with an advection-diffusion PDE subject to appropriate boundary conditions.

% that accounts for lipid-dependent rates of uptake and offloading. The model differs from the established lipid-structured models mentioned above in that it (i) assumes macrophage lipid content is bounded from above, and (ii) is fundamentally based on a discrete structuring of the macrophage population. Nonetheless, we use a continuum approximation to facilitate analysis of the macrophage lipid distribution in Section \ref{sec: lipid distribution}. Furthermore, the current model implicitly treats the uptake of dead cells (efferocytosis) as a piecemeal process. This assumption is consistent with observations that apoptotic cell uptake \textit{in vivo} is often performed piecemeal and by multiple phagocytes (Taefehshokr et al. \cite{taefehshokr2021rab}). 

The remainder of the paper is structured as follows. Section \ref{sec: Model development} describes the model development, including the derivation of an ODE subsystem, non-dimensionalisation and treatment of model parameters. Section \ref{sec: Results} contains the results of our model analysis. We present time-dependent numerical solutions and a steady state analysis of the ODE subsystem and macrophage lipid distribution. Finally, we summarise and discuss the implications of our results in Section \ref{sec: Discussion}, and present concluding remarks in Section \ref{sec: conclusions}.

%----------------------------------------------------------------------------------------------

\section{Model development} \label{sec: Model development}

In this section we present a lipid-structured model for MDM populations in early atherosclerotic lesions. We begin by outlining our assumptions and formulating the model as a coupled system of ODEs. We then show that the model admits a closed subsystem that can be solved and analysed independently of the full model. Finally we non-dimensionalise the model and outline our treatment of the dimensionless parameters.

%---------------------------------------------------------------------------------------------------

\subsection{Definitions and model statement}

We assume for simplicity that macrophages take up and offload lipid in increments of mass $\Delta a$, and decompose the MDM population into a series of compartments according to lipid load. Specifically, we let $m_j(t) \geq 0$ be the concentration of lesion MDMs with lipid content $a_0 + j \Delta a$ at time $t \geq 0$. Here $a_0$ is the mass of endogenous lipid that macrophages contain due to their membrane structure. The index $j$ runs over integer values $j = 0, 1, \dots, J$, where $J>0$ defines the macrophage lipid capacity. It follows that the maximal macrophage lipid content, $\kappa$, is given by $\kappa = a_0 + J \Delta a$. Finally, we denote the concentration of extracellular lipid by $L(t)$, and the total lipid capacity of the HDL particles in the lesion by $H(t)$, with $H(t)$ proportional to the concentration of unloaded HDL particles by a factor $\Delta a$.

We propose that the MDM population evolves according to the following ODEs:
 \begin{align}
    \frac{d m_0}{dt} &= \sigma_M \frac{L}{L_c + L}  - k_L L m_0 + k_H H \Big( \frac{1}{J} \Big) m_1 - (\beta + \gamma) m_0, \label{eqn: m0}\\
\begin{split} \label{eqn: mj}
    \frac{d m_j}{dt} &= \, \, k_L L \Big[ \Big( 1 - \frac{j-1}{J} \Big) m_{j-1} - \Big( 1 - \frac{j}{J} \Big) m_j \Big] \\
    &\quad + k_H H \Big[ \Big( \frac{j+1}{J} \Big) m_{j+1} - \Big( \frac{j}{J} \Big) m_j \Big] - (\beta + \gamma) m_j, \quad 1 \leq j \leq J-1, 
    \end{split} \\
    \frac{d m_J}{dt} &= k_L L \Big( \frac{1}{J} \Big) m_{J-1} - k_H H m_J - (\beta + \gamma) m_J. \label{eqn: mJ}
\end{align}
Equations \eqref{eqn: m0}-\eqref{eqn: mJ} account for lipid uptake and offloading via the Law of Mass Action. Lipid uptake is modelled via the reaction:
\begin{align}
    &m_j(t) + L(t) \xrightarrow[]{k_L (1 - \frac{j}{J})} m_{j+1}(t), & &j = 0, 1, \dots, J-1.  \label{eqn: uptake reaction}
\end{align}
Here we assume that the rate of lipid uptake decreases linearly with lipid content; it takes its maximal value $k_L$ when $j = 0$ and decreases to zero at $j = J$, so that macrophages cannot exceed $J$ units of ingested lipid. Our treatment of lipid uptake is consistent with current understanding that macrophages have a finite capacity for phagocytosis, and that phagocytosis is mediated by intracellular signalling \cite{zent2017maxed}. Lipid offloading is modelled via the reaction:
\begin{align}
    &m_j(t) + H(t)  \xrightarrow[]{k_H (\frac{j}{J})} m_{j-1}(t), & &j = 1, 2, \dots, J, \label{eqn: offload reaction}
\end{align}
so that the rate constant for offloading increases linearly from zero at $j = 0$ to its maximal value $k_H H$ at $j = J$. This formulation is consistent with experimental evidence that intracellular accumulation of cholesterol leads to the up-regulation of transporters (e.g. ABCA1) that mediate the efflux of free cholesterol \cite{remmerie2018macrophages}. 

Equations \eqref{eqn: m0}-\eqref{eqn: mJ} also account for macrophage recruitment, cell death and emigration. We assume that the rate at which macrophages are recruited to the lesion from the bloodstream can be described via a first-order Hill function in $L(t)$. The maximal recruitment rate is $\sigma_M$ and half-maximal recruitment occurs at the value $L = L_c$. We assume that newly recruited macrophages carry only endogenous lipid (corresponding to $j=0$). Therefore, the recruitment source term appears in equation \eqref{eqn: m0} only. We further assume that all lesion macrophages undergo cell death at rate $\beta>0$, and emigrate from the lesion at rate $\gamma>0$.

We propose that the extracellular lipid dynamics are governed by:
\begin{align}
    \frac{dL}{dt} &= \sigma_L + \beta \sum_{j = 0}^J (a_0 + j \Delta a) m_j - k_L \Delta a \, L  \sum_{j = 0}^J \Big( 1 - \frac{j}{J} \Big)m_j - \delta_L L. \label{eqn: L}    
\end{align}
In equation \eqref{eqn: L} we assume that extracellular lipid is sourced from the bloodstream at rate $\sigma_L$, due to LDL particle influx. We assume further that the lipid content of dying macrophages is transferred to the extracellular space, giving rise to the second source term in equation \eqref{eqn: L}. Recall that $\kappa = a_0 + J \Delta a$ is the maximal macrophage lipid content. The third term accounts for lipid removal from the extracellular space by macrophages, at a rate proportional to $k_L$ (see reaction \eqref{eqn: uptake reaction}). Finally, we assume that extracellular lipid is lost to non-MDM interactions (e.g. uptake by the self-renewing population of tissue-resident cells \cite{williams2020limited}) at a constant rate $\delta_L$, which we will assume to be small.

We model the HDL dynamics via the equation:
\begin{align}
    \frac{dH}{dt} &= \sigma_H - k_H  \Delta a \, H \sum_{j = 0}^J \Big( \frac{j}{J} \Big) m_j - \delta_H H. \label{eqn: H}
\end{align}
The ODE \eqref{eqn: H} assumes that HDL lipid capacity is sourced at rate, $\sigma_H$, due to a constant influx of unloaded HDL particles from the bloodstream, and lost to macrophage offloading at a rate proportional to $k_H$ (according to the reaction \eqref{eqn: offload reaction}). The final term accounts for loss of unloaded HDL to non-MDM interactions (e.g. lipid-loading by tissue-resident cells) at the constant rate $\delta_H$, which we will assume to be small.  

\begin{figure}
    \centering
    \includegraphics[width=1.0\textwidth]{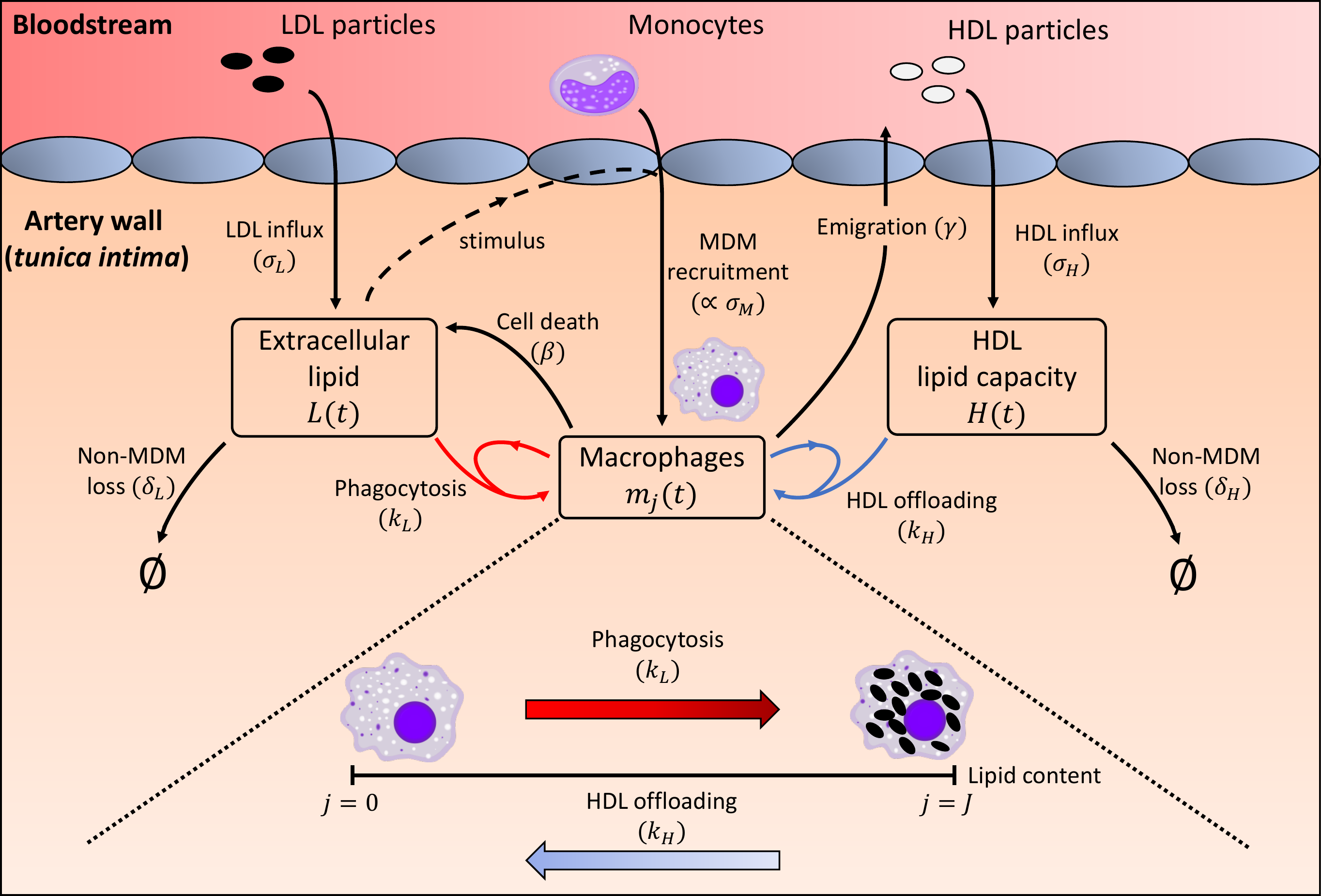}
    \caption{\textbf{Schematic representation of the model \eqref{eqn: m0}-\eqref{eqn: initconds}.} We track extracellular lipid, $L(t)$, the lipid capacity of the HDL particles, $H(t)$, and the number of macrophages, $m_j(t)$, with lipid content $a_0 + j \Delta a$ ($j = 0, 1, \dots, J$). Note the opposing effect of LDL vs HDL particle influx on macrophage lipid content. The LDL influx provides a source of extracellular lipid and hence promotes increases to macrophage lipid load via phagocytosis. By contrast, the entry of HDL promotes decreases in macrophage lipid content via offloading.}
    \label{fig:schematic}
\end{figure}

We close equations \eqref{eqn: m0}-\eqref{eqn: mJ}, \eqref{eqn: L} and \eqref{eqn: H} by imposing the following initial conditions:
\begin{align}
    m_j(0) &= 0, \quad j = 0, 1, \dots, J, & L(0) &= 0, & H(0) &= 0. \label{eqn: initconds}
\end{align}
These equations assume that the lesion is initially devoid of macrophages, extracellular lipid and HDL. We provide a schematic representation of the model in Figure \ref{fig:schematic}. 

%---------------------------------------------------------------------------------------------------

\subsection{An ODE subsystem}

A closed ODE subsystem can be derived from equations \eqref{eqn: m0}-\eqref{eqn: initconds} by considering the total concentration of lesion macrophages, $M(t)$, and the total concentration of intracellular lipid, $A_M(t)$, where,
\begin{align}
    M(t) &:= \sum_{j = 0}^J m_j(t), & A_M(t) &= \sum_{j = 0}^J (a_0 + j \Delta a) m_j(t). \label{eqn: M A_M defs}
\end{align}
Summing equations \eqref{eqn: m0}-\eqref{eqn: mJ}, we find that $M(t)$ evolves according to the equation:
\begin{align}
    \frac{dM}{dt} &= \sigma_M \frac{L}{L_c + L}  - (\beta + \gamma) M.  \label{eqn: M}
\end{align}
Equation \eqref{eqn: M} reflects our assumptions that lesion macrophages are sourced from the bloodstream at rate which is an increasing, saturating function of the extracellular lipid load, $L(t)$, and removed via apoptosis and emigration at constant rates $\beta$ and $\gamma$ respectively. By multiplying each of equations \eqref{eqn: m0}-\eqref{eqn: mJ} by the corresponding index $j$ and then summing, it is straightforward to show that the dynamics of $A_M(t)$ are given by:
\begin{align}
\begin{split} \label{eqn: A_M}
    \frac{d A_M}{dt} &= a_0 \sigma_M \frac{L}{L_c + L}  + \frac{k_L \, \Delta a}{\kappa - a_0} (\kappa M - A_M)L  - \frac{k_H \Delta a}{\kappa - a_0} (A_M - a_0 M) H - (\beta + \gamma ) A_M.
\end{split}
\end{align}
Equation \eqref{eqn: A_M} shows how intracellular lipid levels increase due to the influx of macrophages from the bloodstream and uptake of extracellular lipid. We note that the rate of uptake across the macrophage population is proportional to the quantity $\kappa M(t) - A_M(t)$, which represents the lipid capacity of the macrophage population at time $t$. Intracellular lipid is removed to offloading to HDL, apoptosis and emigration. The rate of lipid offloading across the macrophage population is proportional to $A_M(t) - a_0 M(t)$, which is the total amount of ingested lipid (i.e. non-endogenous lipid) in the macrophage population. 

Using definitions \eqref{eqn: M A_M defs} we simplify equations \eqref{eqn: L} and \eqref{eqn: H} to obtain:
\begin{align}
    \frac{dL}{dt} &= \sigma_L + \beta A_M - \frac{k_L \Delta a}{\kappa - a_0} (\kappa M - A_M) L - \delta_L L, \\
    \frac{dH}{dt} &= \sigma_H -\frac{k_H \Delta a}{\kappa - a_0} (A_M - a_0 M) H - \delta_H H. \label{eqn: H2}
\end{align}
We note that equations \eqref{eqn: M}-\eqref{eqn: H2} define a closed subsystem for $M$, $A_M$, $L$ and $H$, which we solve subject to the following initial conditions:
\begin{align}
    M(0) &= 0, & A_M(0) &= 0, & L(0) &= 0, & H(0) &= 0, \label{eqn: ODE initconds}
\end{align}
which we obtain from equations \eqref{eqn: initconds}.

%---------------------------------------------------------------------------------------------------
\subsection{Non-dimensionalisation}
We recast the model in terms of the following dimensionless variables:
\begin{align}
\begin{aligned}
    \hat{t} &:= \beta t, & \hat{m}_j(\hat{t}) &:= \frac{\beta}{\sigma_M} m_j(t), & \hat{M}(\hat{t}) &:= \frac{\beta}{\sigma_M} M(t), \\
    \hat{L}(\hat{t}) &:= \frac{\beta}{a_0 \sigma_M} L(t), & \hat{A}_M(\hat{t}) &:= \frac{\beta}{a_0 \sigma_M} A_M(t), & \hat{H}(\hat{t}) &:= \frac{\beta}{a_0 \sigma_M} H(t)
\end{aligned} \label{eqn: nondim}
\end{align}
Under this rescaling, time is measured in units of average macrophage lifespan, $\beta^{-1}$, and the macrophage concentrations are normalised with respect to $\frac{\sigma_M}{\beta}$, the maximal concentration of MDMs supplied to the lesion per macrophage lifespan. The lipid concentrations $L(t)$, $A_M(t)$ and the HDL lipid capacity, $H(t)$, are scaled with respect to $a_0 \sigma_M/\beta$, which is the maximal concentration of endogenous lipid entering the lesion per macrophage lifespan.
\begin{table}
\begin{tabular}{|M{3cm}|M{2cm}|l|}
\hline
\textbf{Dimensionless parameter}     & \textbf{Definition}   & \textbf{Physical interpretation} \\ \hline
$\hat{\kappa}$ & $\frac{\kappa}{a_0}$ & \begin{tabular}{@{}l@{}} Ratio of the maximum macrophage lipid content to  \\ endogenous   lipid content \end{tabular} \\ \hline
$J$ & $\frac{\kappa - a_0}{\Delta a}$ & \begin{tabular}{@{}l@{}} Ratio of macrophage lipid capacity to the increment\\ of lipid uptake/offloading  \end{tabular} \\ \hline
$\hat{\sigma}_L$ & $\frac{\sigma_L}{a_0 \sigma_M}$ & \begin{tabular}{@{}l@{}} Influx of LDL lipid relative to maximal influx \\ of endogenous lipid \end{tabular} \\ \hline
$\hat{\sigma}_H$ & $\frac{\sigma_H}{a_0 \sigma_M}$ & \begin{tabular}{@{}l@{}} Influx of HDL capacity relative to maximal influx \\ of endogenous lipid \end{tabular} \\ \hline
$\hat{k}_L$ & $\frac{k_L \Delta a \, a_0 \sigma_M}{\beta^2 (\kappa - a_0)}$ & \begin{tabular}{@{}l@{}} Dimensionless rate of lipid uptake per unit of \\ extracellular lipid \end{tabular} \\ \hline
$\hat{k}_H$ & $\frac{k_H \Delta a \, a_0 \sigma_M}{\beta^2 (\kappa - a_0)}$ & \begin{tabular}{@{}l@{}} Dimensionless rate of lipid offloading per unit of \\ HDL capacity \end{tabular} \\ \hline
$\hat{L}_{c}$ & $\frac{L_{c} \beta}{a_0 \sigma_M}$ & \begin{tabular}{@{}l@{}} Value of $\hat{L}$ for which macrophage recruitment is \\ half-maximal \end{tabular} \\ \hline
$\hat{\gamma}$ & $\frac{\gamma}{\beta}$ & \begin{tabular}{@{}l@{}} Dimensionless rate of emigration \\ \hline
\end{tabular} \\ \hline
$\hat{\delta}_L$ & $\frac{\delta_L}{\beta}$ & \begin{tabular}{@{}l@{}} Dimensionless rate of extracellular lipid loss \\
to non-MDM interactions \\ \hline
\end{tabular} \\ \hline
$\hat{\delta}_H$ & $\frac{\delta_H}{\beta}$ & \begin{tabular}{@{}l@{}} Dimensionless rate of unloaded HDL loss \\ to non-MDM interactions
\\ \hline
\end{tabular} \\ \hline
\end{tabular}
\caption{The dimensionless parameters that appear in equations \eqref{eqn: m0 nondim}-\eqref{eqn: L nondim}.}
\label{tab:parameters}
\end{table}

We also define a number of dimensionless parameters in Table \ref{tab:parameters}. The first two parameters, $\hat{\kappa}$ and $J$, are lipid ratios. The remaining parameters pertain to the model kinetics:  $\hat{\sigma}_L$ and $\hat{\sigma}_H$, determine the rates at which LDL and HDL respectively enter the lesion; $\hat{k}_L$ is the dimensionless phagocytosis rate per unit extracellular lipid; $\hat{k}_H$ is the dimensionless offloading rate per unit HDL capacity; $\hat{L}_c$ is the value of $\hat{L}(\hat{t})$ at which macrophage recruitment is half-maximal; $\hat{\gamma}$ is the dimensionless emigration rate; $\hat{\delta}_L$ is the dimensionless rate of extracellular lipid loss to non-MDM effects; and $\hat{\delta}_H$ determines the rate of unloaded HDL loss to non-MDM effects.

% The first two parameters, $\hat{\sigma}_L$ and $\hat{\sigma}_H$, determine the influx of LDL and HDL into the lesion respectively. The next six parameters pertain to the model kinetics: $\hat{k}_L$ is the dimensionless phagocytosis rate per unit extracellular lipid; $\hat{k}_H$ is the dimensionless offloading rate per unit HDL capacity; $\hat{L}_c$ is the value of $\hat{L}(\hat{t})$ for which macrophage recruitment is half-maximal; $\hat{\gamma}$ is the dimensionless emigration rate; $\hat{\delta}_L$ is the dimensionless rate of extracellular lipid loss to non-MDM interactions; and $\hat{\delta}_H$ determines the rate of unloaded HDL loss to non-MDM interactions. The final two parameters, $\hat{\kappa}$ and $J$, are lipid ratios.

By applying the scaling \eqref{eqn: nondim} and the definitions of Table \ref{tab:parameters}, and dropping the hats for notational convenience, we obtain the following dimensionless ODEs for the macrophage dynamics:
\begin{align}
    \frac{d m_0}{dt} &= \frac{L}{L_c + L} - k_L J L m_0 + k_H H  m_1 - (1 + \gamma) m_0, \label{eqn: m0 nondim}\\
\begin{split}
    \frac{d m_j}{dt} &= k_L L \big[ (J - j + 1) m_{j-1} - (J-j) m_j \big] \\
    &\quad + k_H H \big[ (j+1) m_{j+1} - j m_j \big] - (1 + \gamma)m_j, \quad  1 \leq j \leq J-1, 
\end{split} \label{eqn: mj nondim}\\
\frac{d m_J}{dt} &= k_L L m_{J-1} - k_H H J m_J - (1 + \gamma)m_J, \label{eqn: mJ nondim}
\end{align}
and the following closed subsystem for $M$, $A_M$, $L$ and $H$:
\begin{align}
    \frac{dM}{dt} &= \frac{L}{L_c + L} - (1+\gamma) M, \label{eqn: M nondim} \\
\begin{split}
    \frac{d A_M}{dt} &= \frac{L}{L_c + L} + k_L (\kappa M - A_M) L  - k_H (A_M -M)H - (1+\gamma) A_M,
\end{split} \label{eqn: A_M nondim} \\
    \frac{dL}{dt} &= \sigma_L + A_M - k_L (\kappa M - A_M) L - \delta_L L,  \label{eqn: L nondim} \\
    \frac{dH}{dt} &= \sigma_H - k_H (A_M - M)H - \delta_H H. \label{eqn: H nondim}
\end{align}
The initial conditions used to close equations \eqref{eqn: m0 nondim}-\eqref{eqn: H nondim} are unchanged and stated here for completeness:
\begin{align}
    m_j(0) &= 0, \quad j = 0, 1, \dots, J, & M(0) = A_M(0) &= L(0) = H(0) = 0. \label{eqn: initconds nondim}
\end{align}
% \noindent \textbf{Note:} 
In their dimensionless forms, $M(t)$ and $A_M(t)$ are related to the functions $m_j(t)$ via the summations:
\begin{align}
    M(t) &= \sum_{j=0}^J m_j(t), & A_M(t) &= M(t) + \frac{\kappa - 1}{J}\sum_{j=0}^J j m_j(t). \label{eqn: M A_M sum nondim}
\end{align}

%---------------------------------------------------------------------------------------------------

\subsection{Model parameter estimation}

In what follows, unless stated otherwise, we use the following dimensionless parameter values in our numerical simulations:
\begin{align}
     \hat{\kappa} &= 30, & J &= 100, & \hat{k}_L = \hat{k}_H &= 1,  & \hat{\gamma} &= 0.25, & \hat{L}_c &= 1, & \hat{\delta}_L = \hat{\delta}_H &= 0.01. \label{eqn: parvals}
\end{align}
The influx ratios $\hat{\sigma}_L$ and $\sigma_H$ are varied over a range of positive values. We briefly justify these choices below.

The estimate for $\hat{\kappa} = \kappa/a_0$ is based on the experimental results presented in \cite{ford2019efferocytosis}. Ford \textit{et al.} measured the rate at which macrophages uptake endogenous lipid from apoptotic cells and found that, even when extracellular lipid was in abundance, macrophages did not attain a lipid load exceeding $\sim 30 a_0$.

The average unit of lipid exchange, $\Delta a$, is likely to be much smaller than the macrophage capacity, $\kappa - a_0$. Indeed, the largest increment of lipid exchange is likely to be due to apoptotic cell uptake (efferocytosis) or necrotic cell uptake, processes which are often reported to occur via piecemeal uptake or via nibbling respectively \cite{taefehshokr2021rab, westman2020phagocytosis}. Hence, a possible upper bound for $\Delta a$ in early atherosclerotic lesions is the endogenous lipid content: $\Delta a \lesssim a_0$. In light of our estimate for $\kappa$, it follows that $J^{-1} = \frac{\Delta a}{\kappa - a_0} \lesssim 0.03$.

The estimate for $\hat{\gamma} = \gamma/\beta$ is based on a macrophage death rate of $\beta = 0.05$ h$^{-1}$ and an emigration rate of $\gamma = 0.013$ h$^{-1}$. The stated death rate is the half-maximal value of the apoptosis rate reported in \cite{thon2018quantitative}, where the authors fit an ODE model to \textit{in vitro} data for macrophage lipid uptake. Thon \textit{et al.} assume that the death rate increases with lipid content (by contrast we assume a constant death rate). Since our model distinguished between newly recruited and lipid-laden macrophages, we use the half-maximal rate rather than the maximal rate. The stated emigration rate is intermediate between the 12.6$\%$ transmigration rate reported in \cite{angelovich2017quantification} and the 20h residence time reported in \cite{ghattas2013monocytes}.

The remaining parameters are difficult to estimate due to the current lack of quantitative \textit{in vivo} data. In the absence of such data, we fix $\hat{k}_L = \hat{k}_H = 1$. These values ensure that macrophage phagocytosis and offloading are comparable and occur on the same timescale as the macrophage lifespan. We also fix $\delta_L = \delta_H = 0.01$ so that loss of extracellular lipid and unloaded HDL to non-MDM interactions is small relative to MDM phagocytosis and offloading. This assumption is consistent with observations that the number of tissue-resident cells in atherosclerotic plaques is dominated by the number of recruited MDMs \cite{williams2020limited}. Finally, we set $\hat{L}_c = 1$ so MDM recruitment also occurs on the $t = \mathcal{O}(1)$ macrophage lifespan timescale. The LDL and HDL influx ratios, $\hat{\sigma}_L$ and $\hat{\sigma}_H$, are likely to vary significantly between individuals since blood cholesterol levels depend sensitively on lifestyle factors such as diet \cite{bruckert2011lowering}. Nonetheless, the estimates in \cite{thon2018quantitative} indicate that $\hat{\sigma}_L$ and $\sigma_H$ are likely to be $\mathcal{O}(1)$ quantities.

%---------------------------------------------------------------------------------------------------
%---------------------------------------------------------------------------------------------------

\section{Results} \label{sec: Results}

In this section we explore the behaviour of equations \eqref{eqn: m0 nondim}-\eqref{eqn: initconds nondim}. We present time-dependent numerical solutions of the full model in Subsection \ref{sec: dynamics}. We then conduct a steady state analysis of the subsystem \eqref{eqn: M nondim}-\eqref{eqn: L nondim} in Subsection \ref{sec: ODE steady state}. Finally, in Subsection \ref{sec: lipid distribution}, we characterise the macrophage lipid distribution  by deriving and then exploiting a continuum approximation of equations \eqref{eqn: m0 nondim}-\eqref{eqn: mJ nondim}.

%---------------------------------------------------------------------------------------------------

\subsection{Model dynamics} \label{sec: dynamics}

\begin{figure}
    \centering
    \includegraphics[width=1.0\textwidth]{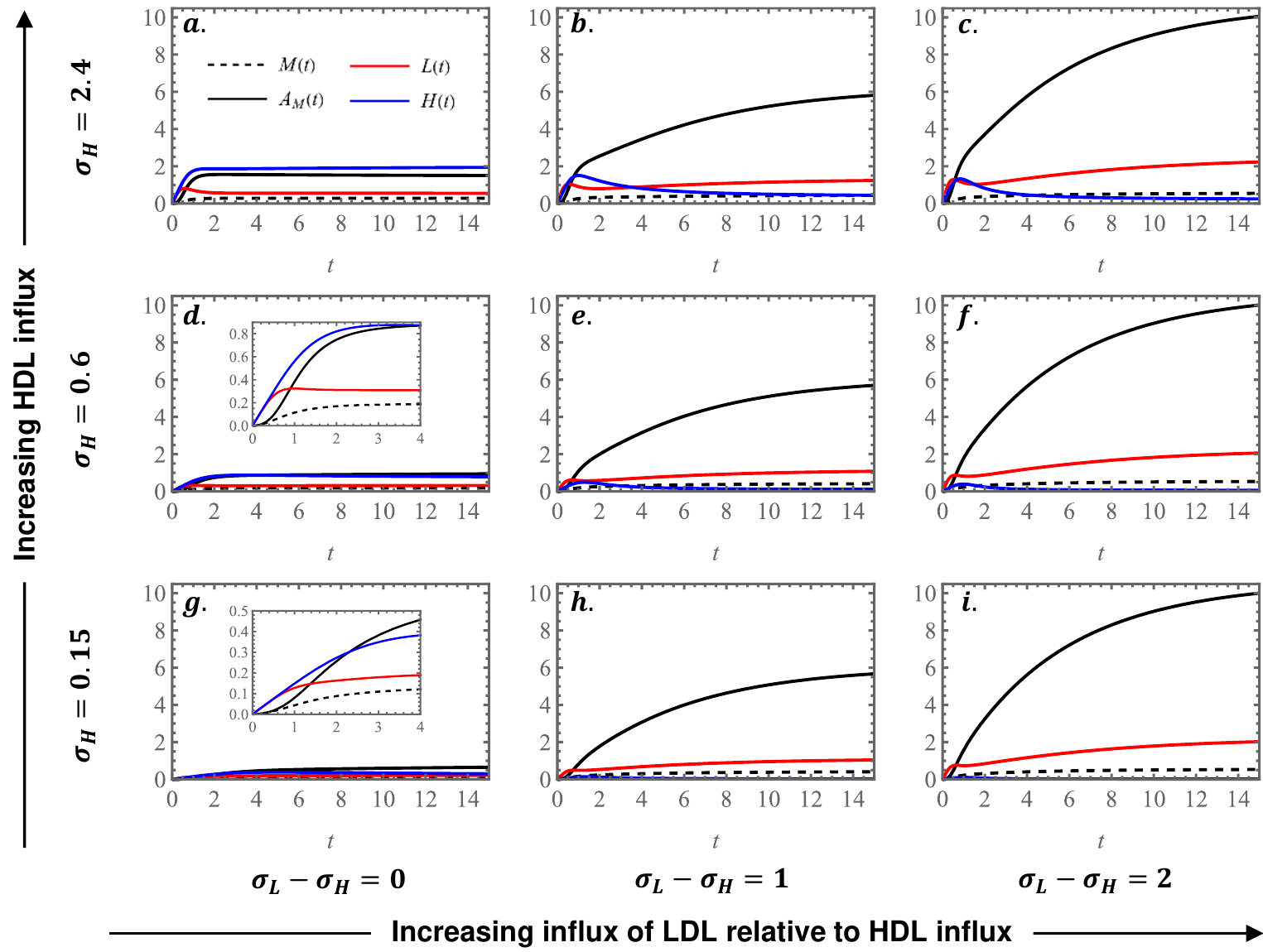}
    \caption{\textbf{Series of plots showing how plaque composition against time for different rates of LDL/HDL entry.} The time dynamics of the macrophage population, $M$, intracellular lipid content, $A_M$, extracellular lipid content, $L$ and HDL capacity for lipid, $H$, are presented for a range of values of the parameters $\sigma_L$ and $\sigma_H$. We consider the cases $\sigma_H = 0.15$, $0.6$ and $2.4$ in the top, middle and bottom rows respectively. The LDL influx, $\sigma_L$, is increased from $\sigma_H$ to $\sigma_H + 1$ and finally $\sigma_H + 2$ from left to right. The remaining parameter values are given in equation \eqref{eqn: parvals}.}
    \label{fig: composition dynamics}
\end{figure}

We use Wolfram Mathematica's \textit{NDSolve} routine to generate numerical solutions across a range of values of the influx parameters $\sigma_L$ and $\sigma_H$. The dynamics of the ODE subsystem \eqref{eqn: M nondim}-\eqref{eqn: initconds nondim} are shown in Figure \ref{fig: composition dynamics}. At early times, $t \lesssim 0.25$ (see insets of Figs. \ref{fig: composition dynamics}d, g), we observe quasi-linear growth in $L(t)$ and $H(t)$, which reflects LDL and HDL infiltrating a lesion that is initially devoid of monocyte-derived macrophages. The initial rise in $L(t)$ drives a rapid increase in $M(t)$ and $A_M(t)$ as macrophages are recruited to the lesion from the blood. The growth in $L(t)$ then slows substantially for $0.5 \lesssim t \lesssim 1$ due to macrophage phagocytosis. We note that this initial wave of phagocytosis is sufficient to cause $L(t)$ to decrease in many of the cases shown, which indicates that the rate of phagocytosis exceeds the rate of LDL entry to the tissue during this period. For $1 \lesssim t \lesssim 2$, growth in $H(t)$ slows, indicating that at this stage the macrophage population has ingested enough lipid for offloading to substantially impact the amount of unloaded HDL in the lesion. Indeed for sufficiently large values of  $(\sigma_L - \sigma_H)$ (see cases b,c,e,f,h,i), intracellular lipid loads are sufficiently large that offloading causes $H(t)$ to peak and decrease thereafter. We note also that macrophage cell death gives rise to a secondary increase in $L(t)$ in these cases during this time period; the increase in extracellular lipid due to cell death exceeds the loss due to phagocytosis for sufficiently large values of  $\sigma_L$ relative to $\sigma_H$. Finally, we observe that $M(t)$, $A_M(t)$, $L(t)$ and $H(t)$ evolve to nonzero equilibrium values as $t \rightarrow \infty$.

The numerical solutions suggest that plaque lipid content at equilibrium is more sensitive to the relative influx of LDL to HDL, $(\sigma_L - \sigma_H)$, than to the individual influx rates $\sigma_L$ and $\sigma_H$. Indeed, when $\sigma_L - \sigma_H = 2$ the trajectories of $A_M(t)$ and $L(t)$ are indistinguishable for $t \gtrapprox 4$ (see Fig.\ref{fig: composition dynamics}, cases c,f,i). We note further that larger values of $(\sigma_L - \sigma_H)$ give rise to larger increases in $A_M(t)$ than in $L(t)$ for the parameter values used in Figure \ref{fig: composition dynamics}. That is, increases to the relative influx of LDL to HDL yield greater changes in the macrophage lipid burden than the amount of extracellular lipid for these parameter values. These trends will be explored in more detail in Section \ref{sec: ODE steady state}.

\begin{figure}
    \centering
    \includegraphics[width=1.0\textwidth]{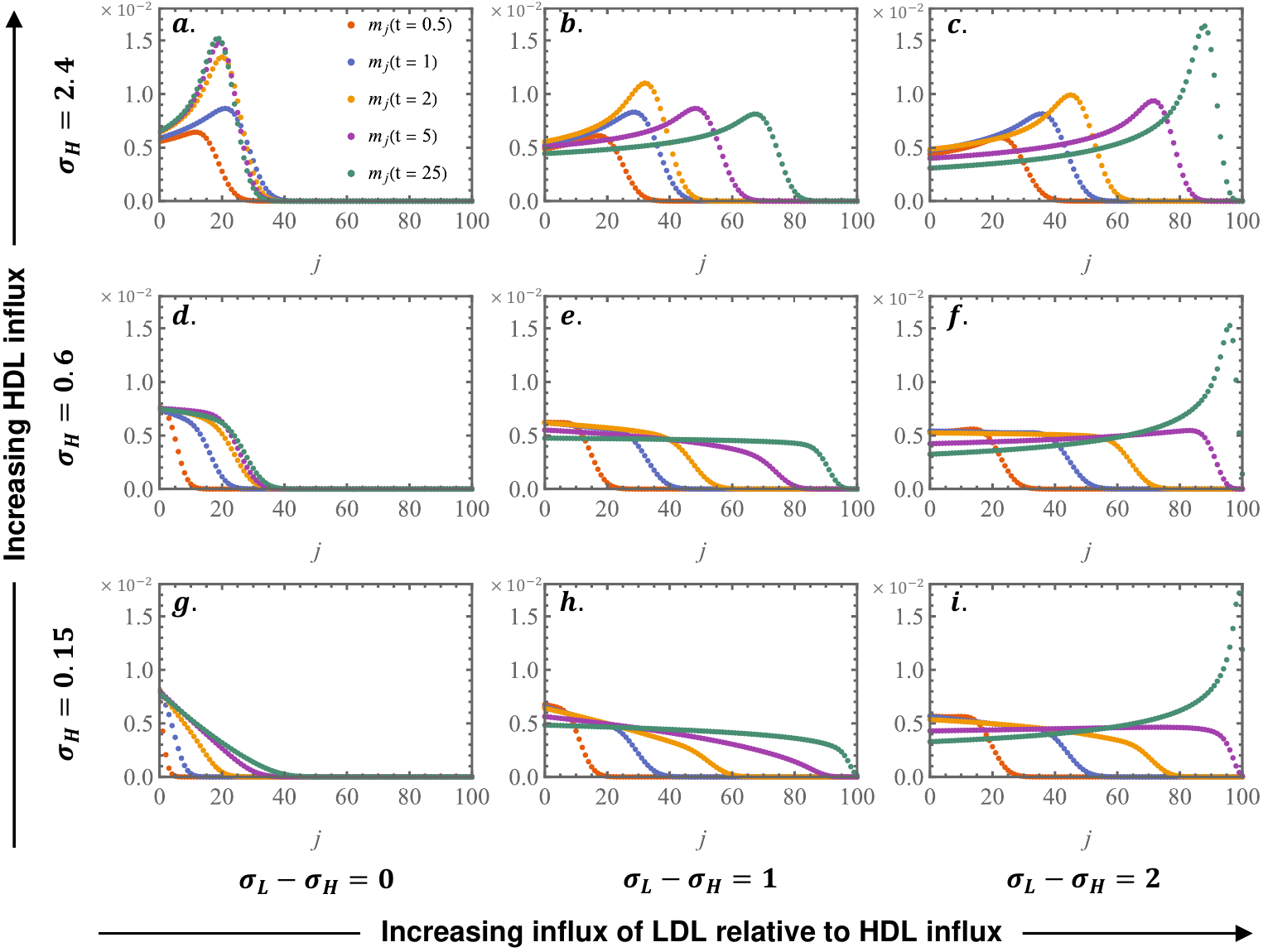}
    \caption{\textbf{Dynamics of the macrophage lipid distribution for different rates of LDL and HDL influx.} The distribution functions, $m_j(t)$, $j = 0, 1, \dots, J$, are plotted against $j$ at the time values $t = 0.5$, $1$, $2$, $5$ and $25$ (approximately steady state). We consider the same values of $\sigma_L$ and $\sigma_H$ as in Figure \ref{fig: composition dynamics}. The remaining parameters are set according to equation \eqref{eqn: parvals}.}
    \label{fig: distribution dynamics}
\end{figure}
The dynamics of the macrophage lipid distribution, $m_j(t)$, $j = 0, 1, \dots, J$, are shown in Figure \ref{fig: distribution dynamics}. We see that, over time,  $m_j(t)$ evolves in a wave-like manner towards a non-zero equilibrium. The wavefront proceeds from $j = 0$ at $t = 0$ to $j=j_c \in (0, J]$ as $t \rightarrow \infty$. We note, however, that the wavefront may change direction at later times (e.g. case a in Fig.\ref{fig: distribution dynamics}). This is likely due to offloading to HDL, given the dominance of $H(t)$ for these parameter values (see Fig.\ref{fig: composition dynamics}a).

The lipid distribution $m_j$ adopts several qualitatively distinct profiles at equilibrium. The equilibrium profile may be monotone decreasing in $j$, and may be convex (as in case g) or concave (cases d and h); it may have a quasi-uniform profile (case e) or a peaked profile (cases a,b,c,f and i). The simulation results presented in Figure \ref{fig: distribution dynamics} suggest that the location of the wavefront at steady state is sensitive to the difference $(\sigma_L - \sigma_H)$. However, the geometric features of the profile (e.g. existence of a local maximum) seem to depend on the individual magnitudes of $\sigma_L$ and $\sigma_H$. For instance, the spiked profile can manifest with low (e.g. case a), intermediate (e.g. case b) or high (e.g. case c) lipid loads. We study the equilibrium lipid distribution in more depth in Section \ref{sec: lipid distribution}.

Before proceeding to the steady state analysis, we pause to define two quantities which help to summarise the model behaviour:
\begin{align}
    \mu(t) &:= \frac{A_M(t)-M(t)}{(\kappa - 1)M(t)}, & f(t) &:= \frac{A_M(t)}{A_M(t) + L(t)}. \label{eqn: mu f defns}
\end{align}
Both variables are normalised such that $0 \leq \mu(t), f(t) \leq 1$. The ratio $\mu(t)$ represents the average macrophage lipid burden, relative to the capacity for ingested lipid. Since
\begin{align}
    \mu = 1 - \frac{k_L (\kappa M - A_M) L}{k_L (\kappa M - M) L},
\end{align}
we note that another interpretation for $\mu$ is the fractional decrease in the phagocytosis rate of the macrophage population due to the lipid load at time $t$. The ratio $f(t)$ is the fraction of lipid in the lesion that is contained within the macrophage population.

\begin{figure}
    \centering
    \includegraphics[width=1.0\textwidth]{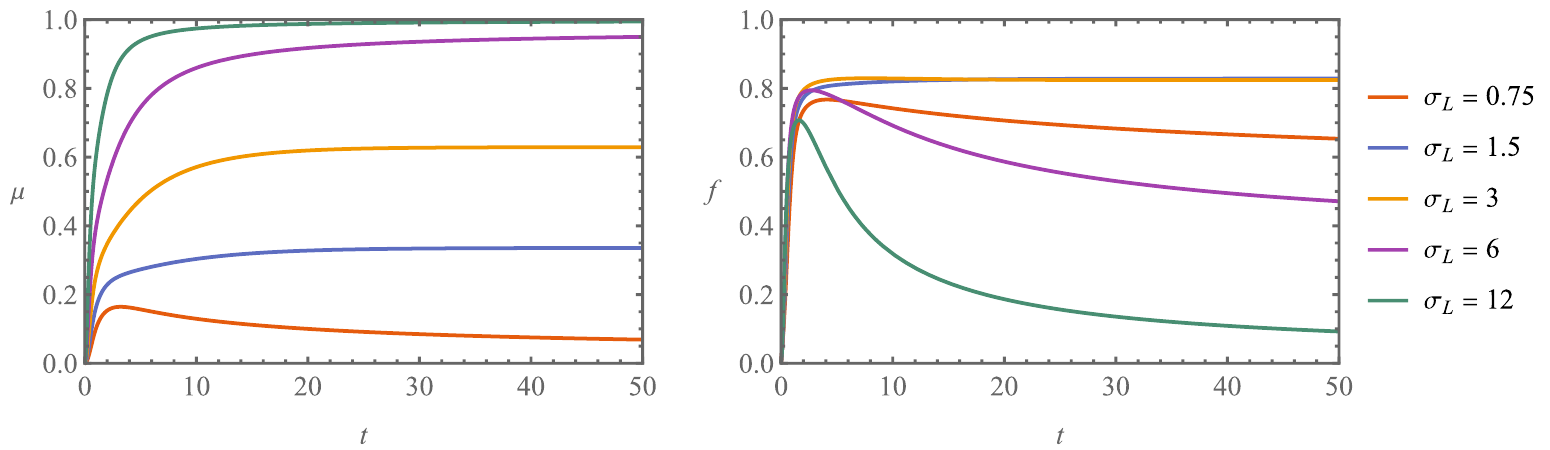}
    \caption{\textbf{Fractional decrease in phagocytosis rate due to lipid accumulation, $\mu(t)$, and proportion of lesion lipid that is intracellular, $f(t)$.}
    The quantities $\mu(t)$ and $f(t)$, defined in equations \eqref{eqn: mu f defns}, are plotted for the cases $\sigma_L = 0.75, 1.5, 3, 6$ and $12$ with $\sigma_H = 1$. We set the remaining parameters according to equation \eqref{eqn: parvals}.}
    \label{fig:summary_stats}
\end{figure}

The dynamics of $\mu(t)$ and $f(t)$ are shown in Figure \ref{fig:summary_stats} for $\sigma_H = 1$ and $\sigma_L = 0.75$, $1.5$, $3$, $6$ and $12$. Both variables tend to non-zero steady state values at long times: $\mu \rightarrow \mu^\star$ and $f \rightarrow f^\star$ as $t \rightarrow \infty$. We observe that as $\sigma_L$ increases, $\mu^\star$ increases towards $\mu^\star \rightarrow 1^{-}$, which corresponds to a plaque with macrophage lipid loads at capacity and, consequently, $100\%$ defective phagocytosis. By contrast, $f^\star$ depends non-monotonically on $\sigma_L$, attaining its largest value, $f^\star \approx 0.8$, when $1.5 \lesssim \sigma_L \lesssim 3$. In order to understand these trends and those described above, we turn now to the steady state analysis.

%---------------------------------------------------------------------------------------------------
    
\subsection{Steady state analysis of subsystem \eqref{eqn: M nondim}-\eqref{eqn: initconds nondim}} \label{sec: ODE steady state}
Let $M^\star$, $A_M^\star$, $L^\star$ and $H^\star$ denote the steady values of the corresponding dependent variables. Expressions for these quantities can be found by setting the time derivatives to zero in equations \eqref{eqn: M nondim}-\eqref{eqn: L nondim}. We find that $M^\star$, $A_M^\star$ and $H^\star$ can be expressed in term of $L^\star$:
\begin{align}
\begin{aligned}
        M^\star &= \frac{L^\star}{(1 + \gamma)(L_c + L^\star)}, &    A_M^\star &= \frac{k_L \kappa (L^\star)^2-\sigma_L (1+\gamma)(L^\star + L_c)}{(1 + \gamma)(1+k_L L^\star)(L^\star + L_c)},   \\
        H^\star &= \frac{\sigma_H}{\delta_H + k_H (A_M^\star - M^\star)},
\end{aligned} \label{eqn: steadysol general}
\end{align}
and that $L^\star$ solves a quintic equation:
\begin{align}
   c_5 (L^\star)^5 +  c_4 (L^\star)^4+  c_3 (L^\star)^3 +  c_2 (L^\star)^2 +  c_1 L^\star + c_0 = 0. \label{eqn: Lstar quintic}
\end{align}
The coefficients $c_i$, $i = 1, \dots, 5$ are complicated combinations of the model parameters. Their expressions are given in Appendix A. We note that physically realistic steady state solutions must satisfy $A_M^\star \geq M^\star$ so that macrophage lipid content is not less than the endogenous lipid content. Re-writing this condition using the expressions for $A_M^\star$ and $M^\star$ in equations \eqref{eqn: steadysol general} supplies the following inequality for $L^\star$:
\begin{align}
\begin{aligned}
    L^\star &\geq  \, \frac{1}{2\big(k_L (\kappa - 1) + \delta_L (1+\gamma)\big)} \bigg[ \big(1 - (1+\gamma)(\sigma_L - \delta_L L_c)\big) \\
     &\qquad \qquad+ \sqrt{\big(1 - (1+\gamma)(\sigma_L - \delta_L L_c)\big)^2 - 4\big(k_L (\kappa - 1) + \delta_L (1+\gamma)\big) (1+\gamma)\sigma_L L_c} \, \bigg] \\
     &=: L^\star_{\text{min}}. \label{eqn: Lmin}
\end{aligned}
\end{align}
Hence, we seek positive solutions of the quintic \eqref{eqn: Lstar quintic} which satisfy inequality \eqref{eqn: Lmin}.

\begin{figure}
    \centering
    \includegraphics[width=1.0\textwidth]{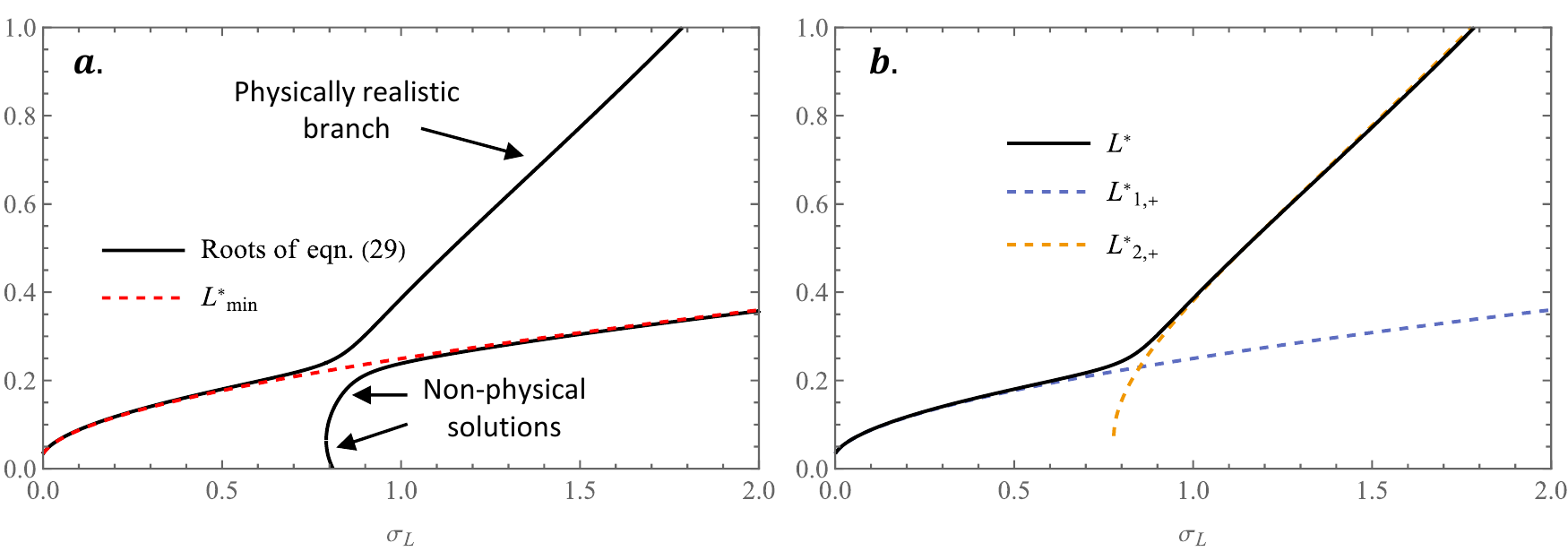}
    \caption{\textbf{Identifying and approximating the physical solution, $L^\star$, of the polynomial \eqref{eqn: Lstar quintic}.} The left plot shows the positive solutions of equation \eqref{eqn: Lstar quintic} as the LDL influx, $\sigma_L$, is varied. There is only one physically realistic solution which satisfies $L^\star \geq L^\star_{\text{min}}$. The right plot shows how the left and right segments of this physical solution are approximated by the asymptotic solutions, $L^\star_{1,+}$ and $L^\star_{2,+}$, respectively. We use $\sigma_H = 1$ and set the remaining parameter values according to equation \eqref{eqn: parvals}.}
    \label{fig:L polyplot}
\end{figure}

We used the Wolfram Mathematica routine \textit{NSolve} to determine the positive solutions of equation \eqref{eqn: Lstar quintic} as $\sigma_L$ varies. The results, presented in Figure \ref{fig:L polyplot}a, show that for fixed values of $\sigma_L$ polynomial \eqref{eqn: Lstar quintic} admits a unique physically realistic solution (with $L^\star \geq L^\star_{\text{min}}$). Further, $L^\star$ increases continuously with $\sigma_L$ and the solution branch can be decomposed into two main segments. On the left segment ($0 \leq \sigma_L \lessapprox 0.8$ in Fig.\ref{fig:L polyplot}) $L^\star$ increases slowly with $\sigma_L$ and is well-approximated by $L^\star_{\text{min}}$. This regime corresponds to an HDL-dominant plaque where macrophage lipid loads are minimal ($A_M^\star \approx M^\star$) due to efficient offloading. In this regime, increasing LDL influx, $\sigma_L$, has little effect on extracellular lipid accumulation at equilibrium because macrophages rapidly offload the lipid they ingest and, therefore, only contribute their endogenous lipid content to the extracellular environment upon death. For large values of $\sigma_L$ ($\sigma_L \gtrapprox 0.8$ in Fig.\ref{fig:L polyplot}) $L^\star$ increases more rapidly with $\sigma_L$. For these plaques the LDL influx, $\sigma_L$, is sufficiently large to impact the quantity of unloaded HDL in the lesion at steady state.

While a closed-form expressions for $L^\star$ does not exist in full generality, we can approximate the individual segments of $L^\star$ via asymptotic analysis. Specifically, we seek approximate solutions to the polynomial \eqref{eqn: Lstar quintic} in the limit $\delta_L \sim \delta_H \ll 1$. Biologically, this corresponds to a regime in which loss of extracellular lipid and unloaded HDL to non-MDM interactions is negligible. It is possible to show that at leading order (see Appendix A), the coefficients scale as: $c_0, \dots, c_4 = \mathcal{O}(1)$ and $c_5 = \mathcal{O}(\delta_L, \delta_H)$, so that equation \eqref{eqn: Lstar quintic} can be interpreted as a singular perturbation problem. We first consider the ``outer" problem by setting $\delta_L, \delta_H \rightarrow 0$ in equation \eqref{eqn: Lstar quintic}. By factorising the resulting quartic polynomial into two quadratic equations for $L^\star$, we arrive at the following solutions:
\begin{align}
    L^\star_{1,\pm} &:= \frac{\gamma  \sigma _L+\sigma _L+1 \pm \sqrt{4 (\gamma +1) (\kappa -1) L_c k_L \sigma _L+(\gamma  \sigma _L+\sigma _L+1){}^2}}{2 (\kappa -1) k_L}, \label{eqn: L1 asymp}\\
    L^\star_{2,\pm} &:= \frac{-b_1 \pm \sqrt{b_1^2 - 4b_0b_2}}{2b_2}, \label{eqn: L2 asymp}
\end{align}
where the constants $b_0, b_1$ and $b_2$ of $L^\star_{2,\pm}$ are given by:
\begin{align}
\begin{aligned}
    b_0 &:=  L_c (1 + \gamma) \big[ (1+\gamma)\sigma_L - \sigma_H \big], \\
    b_1 &:= (\gamma +1) \big[1 + k_L L_c (\sigma_L - \sigma_H) + (1+\gamma)\sigma_L - \sigma_H  \big], \\
    b_2 &:= k_L \big[(1+\gamma)(1+\sigma_L - \sigma_H) - \gamma \kappa \big]. \label{eqn: L2 coeffs}
\end{aligned}
\end{align}
The remaining root is negative and scales as $L^\star_0 = \mathcal{O}(\delta_L^{-1},\delta_H^{-1})$. 

In Figure \ref{fig:L polyplot}b we show how the solutions $L^\star_{1,+}$ and $L^\star_{2,+}$ change as $\sigma_L$ varies. We note that these expressions are good approximations for the left and right segments of $L^\star$ respectively. We note further that $L^\star_{\text{min}}$ reduces to $L^\star_{1,+}$ in the limit $\delta_L \rightarrow 0$, which indicates that $L^\star_{1,+}$ corresponds to a regime where $A_M^\star =  M^\star + \mathcal{O}(\delta_L, \delta_H)$. Substituting this result into equations \eqref{eqn: steadysol general} shows that $H^\star = \mathcal{O}(\delta_L^{-1}, \delta_H^{-1})$ at leading order, confirming that the slower increase with $\sigma_L$ on the left branch $L^\star_{1,+}$ is due to rapid offloading to HDL. The expression for $L^\star_{2,+}$ is more complicated and difficult to interpret. However, we note that the coefficients \eqref{eqn: L2 coeffs} which define $L^\star_{2,+}$ only depend on the influx parameters $\sigma_L$ and $\sigma_H$ through the combinations $\sigma_L - \sigma_H$ and $(1+\gamma) \sigma_L - \sigma_H$. This reinforces the notion that $L^\star$ is sensitive to the relative influx of LDL and HDL.

\begin{figure}
    \centering
    \includegraphics[width=1.0\textwidth]{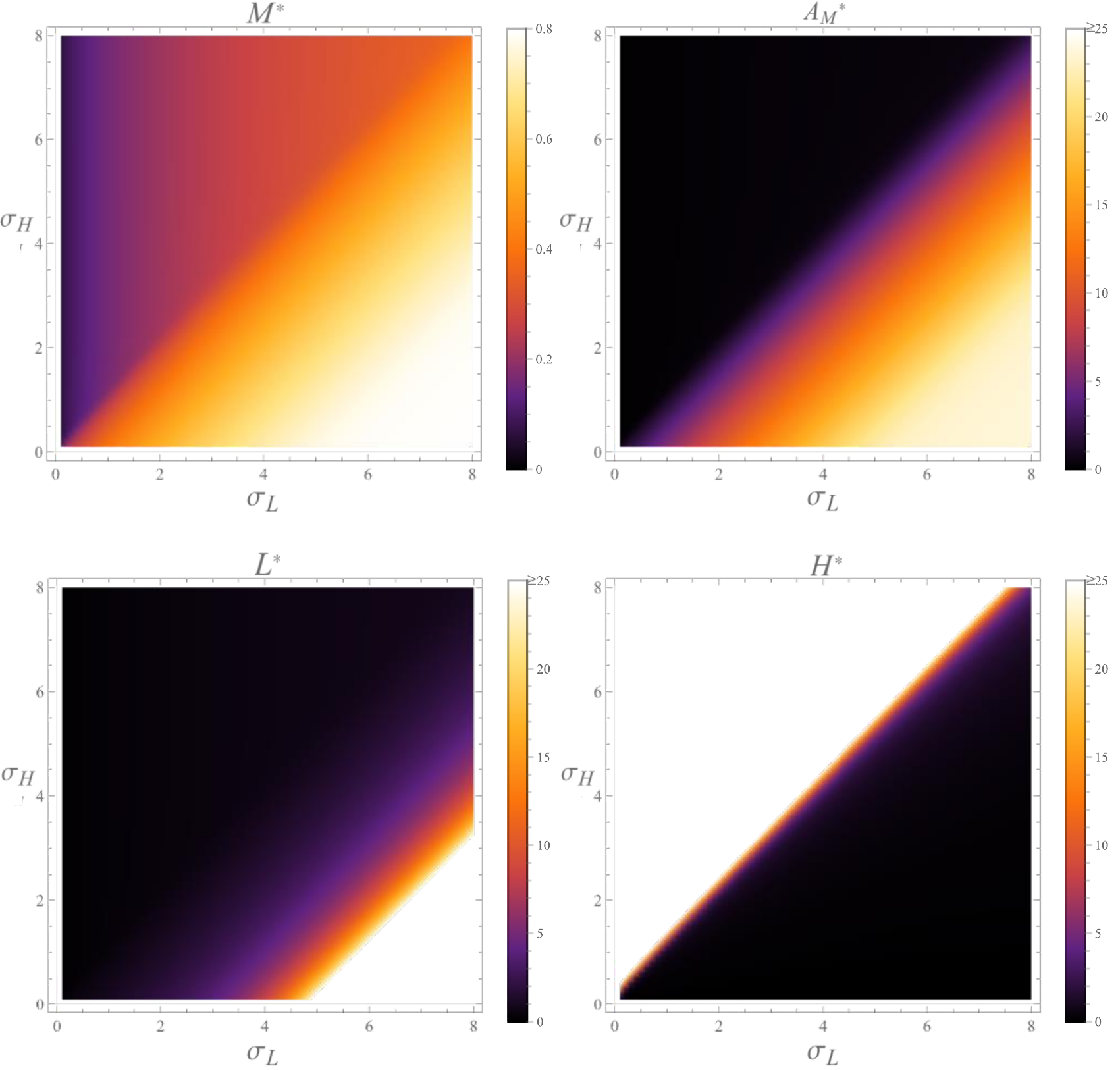}
    \caption{\textbf{Plaque composition is sensitive to the influxes of LDL ($\sigma_L$) and HDL ($\sigma_H$).} The steady state values $M^\star$, $A_M^\star$, $L^\star$ and $H^\star$ are plotted against $\sigma_L$ and $\sigma_H$. Note that the approximately linear contours, which feature in all four plots, indicate that the dependent variables are sensitive to the difference $\sigma_L - \sigma_H$, indicating that plaque composition is largely determined by the relative influx of LDL lipid to HDL capacity. Parameter values are set according to equations \eqref{eqn: parvals}.}
    \label{fig:steady contours}
\end{figure}

In Figure \ref{fig:steady contours}, we use the expressions \eqref{eqn: steadysol general} with the numerical solution for $L^\star$ in equation \eqref{eqn: Lstar quintic} to show how plaque composition at equilibrium changes as both $\sigma_L$ and $\sigma_H$ vary. We find that $M^\star$, $A_M^\star$ and $L^\star$ each increase with $\sigma_L$ and decrease with $\sigma_H$. By contrast, $H^\star$ decreases with $\sigma_L$ and increases with $\sigma_H$. We note that the approximately linear contours, which feature in all four plots, indicate that the dependent variables are sensitive to the difference $\sigma_L - \sigma_H$. This suggests that the relative influx of LDL lipid to HDL capacity not only determines $L^\star$ (as indicated by the asymptotics), but plaque composition more generally. When $\sigma_L -\sigma_H \lesssim 0$ (top left in plots of Fig.\ref{fig:steady contours}), the plaque is HDL-dominant ($H^\star > 25$) with low quantities of lipid ($A_M^\star$, $L^\star$) and a small macrophage population ($M^\star$). For $0 \lesssim \sigma_L - \sigma_H \lesssim 5$ the plaque contains intermediate amounts of macrophages, lipid and HDL capacity. Finally, when $\sigma_L - \sigma_H \gtrsim 5$ the plaque is dominated by extracellular lipid ($L^\star > 25$). In this regime, HDL capacity is exhausted ($H^\star \approx 0$) and macrophage lipid loads are near capacity ($A_M^\star \approx \kappa M^\star$) so that increases to the LDL lipid influx relative to HDL capacity, $(\sigma_L - \sigma_H)$, directly increase the amount of extracellular lipid at steady state ($L^\star$).

\begin{figure}
    \centering
    \includegraphics[width=1.0\textwidth]{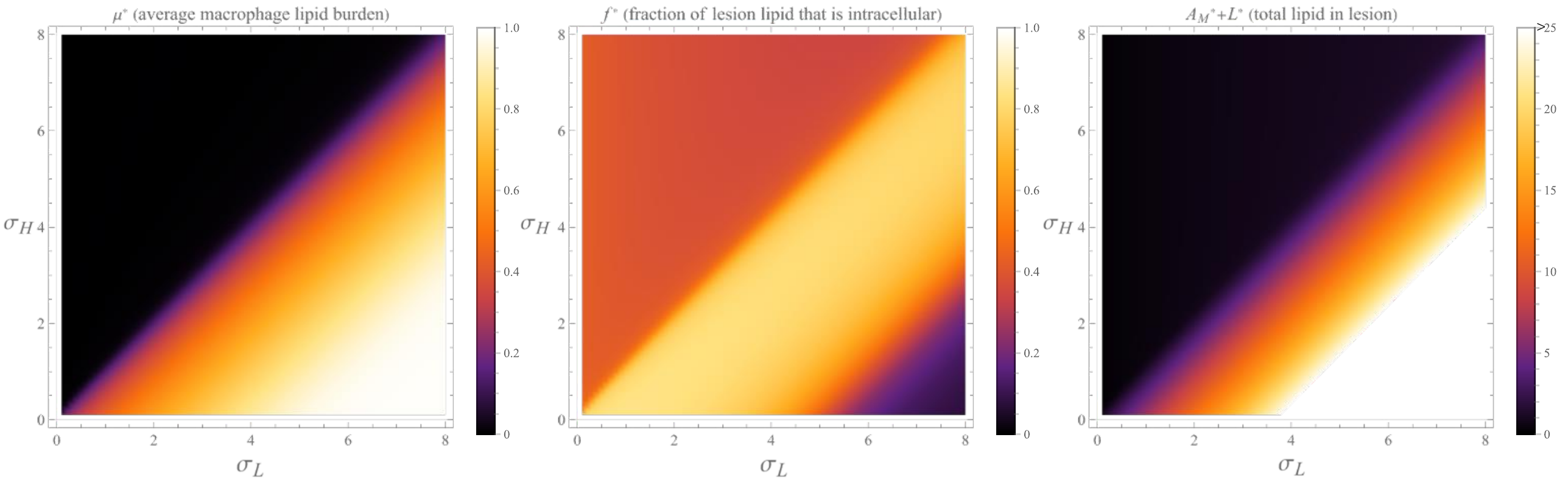}
    \caption{\textbf{Equilibrium plaque lipid content as the LDL and HDL influxes are varied.} We plot the normalised quantities $\mu^\star$ and $f^\star$, defined by equations \eqref{eqn: mu f defns}, and the total lesion lipid, $A_M^\star + L^\star$ against $\sigma_L$ and $\sigma_H$. Note how the fraction of lipid that is intracellular, $f^\star$, is approximately constant in the band $0 \lesssim \sigma_L - \sigma_H \lesssim 5$, in which average macrophage lipid burden, $\mu^\star$, and total lesion lipid, $A_M^\star + L^\star$, both vary substantially. The parameter values are set according to equations \eqref{eqn: parvals}.}
    \label{fig:summary steady}
\end{figure}

We further explore plaque composition in Figure \ref{fig:summary steady} where we focus on $\mu^\star$, the average macrophage lipid burden), $f^\star$, the fraction of lesion lipid that is intracellular and $A_M^\star + L^\star$,  the total lesion lipid, as $\sigma_L$ and $\sigma_H$ vary. In contrast to the trends of $\mu^\star$ and $A_M^\star + L^\star$ (which both increase with $\sigma_L$ and decrease with $\sigma_H$), the fraction $f^\star$ exhibits non-monotonic dependence on both $\sigma_L$ and $\sigma_H$. It follows that $f^\star$ may be a poor metric for plaque progression since it can take the same value when plaque lipid content is low or high (e.g. $f^\star \approx 0.4$ at $(\sigma_L, \sigma_H) = (2,4)$ or $(6,2)$). We note further that $f^\star$ is approximately constant for $0 \lesssim \sigma_L - \sigma_H \lesssim 5$, while $\mu^\star$ and $A_M^\star + L^\star$ vary substantially in this interval. This behaviour probably occurs because, in this region of parameter space, most of the lipid is contained in the macrophage population  (compare plots for $A_M^\star$ and $L^\star$ in Fig.\ref{fig:steady contours}), and so the primary source of extracellular lipid is cell death. Since the amount of lipid entering the extracellular environment via cell death is exactly the lipid content of the dying macrophages, cell death acts to equalise any changes in the amounts of intracellular and extracellular lipid. Finally, the plots confirm that the lesion is dominated by extracellular lipid in the region $\sigma_L -\sigma_H \gtrsim 5$, where macrophage lipid loads are comparable to the maximum lipid capacity. 

\subsection{Macrophage lipid distribution} \label{sec: lipid distribution}

It is not straightforward to analyse the lipid distribution $m_j(t)$, $j = 0, 1, \dots, J$ by directly considering equations \eqref{eqn: m0 nondim}-\eqref{eqn: mJ nondim}. In Appendix B we show that closed-form solutions for the steady state values, $m_j^\star$, $j = 0, 1, \dots J$, can be obtained via the method of generating functions. The resulting steady state solutions read:
\begin{align}
m_j^\star &= M^\star \Bigg( \prod_{\ell=1}^j \frac{a_c (J- \ell + 1)}{(p+1+\ell)} \Bigg) \sum_{n = 0}^{J-j} \frac{(j+1)^{(n)} (j-J)^{(n)}}{(j+2+p)^{(n)}} \frac{a_c^n}{n!}, \quad 0 \leq j \leq J, \label{eqn: m_j formula}
\end{align}
where $x^{(n)} = x (x+1) \cdots (x+n-1)$ is the $n$-th rising factorial of $x \in \mathbb{R}$, and $a_c$ and $p$ are the parameter groupings:
\begin{align}
    a_c &:= \frac{k_L L^\star}{k_L L^\star + k_H H}, & p &:= \frac{1+\gamma}{k_L L^\star + k_H H^\star} - 1.  
\end{align}
Although the solution \eqref{eqn: m_j formula} is exact, the complicated dependence of $m_j^\star$ on the index $j$ and the model parameters makes it difficult to analyse the geometric features of the distribution (e.g. the curvature, existence of a local maximum). We show below how this information can be readily extracted from a continuum approximation of equations \eqref{eqn: m0 nondim}-\eqref{eqn: mJ nondim}. 

\subsubsection{Derivation of continuum approximation} \label{sec: continuum}

We identify the (dimensionless) discrete lipid distribution $m_j(t)$, $j = 0, 1, \dots J$, with the function $m(a,t) \geq 0$, which is defined over a continuous structure variable, $a$. Specifically, we make the identifications:
\begin{align}
     a &\sim  \frac{j}{J} & &\text{and} & m(a,t) &\sim J m_j(t),
\end{align}
so that $0 \leq a \leq 1$ and $m(a,t)$ is proportional to the number density of macrophages with lipid content $a_0 + (\kappa - a_0) a$ at time $t$. The structure variable is bounded such that $a = 0$ corresponds to macrophages containing only endogenous lipid and $a = 1$ to macrophages whose lipid load is at maximum capacity. Letting $\epsilon := J^{-1}$, we recast the discrete equations \eqref{eqn: mj nondim} as the following non-local equation:
\begin{align}
\begin{split}
    \epsilon \frac{\partial m}{\partial t} (a,t) &= k_L L  \big[ (1-a+\epsilon) m(a-\epsilon, t) - (1-a) m(a,t) \big] \\
    &\quad + k_H H \big[ (a + \epsilon) m(a + \epsilon, t) - a m(a,t) \big] - (1+\gamma) m(a,t), \label{eqn: m nonlocal}
\end{split}
\end{align}
where $\epsilon \leq a \leq 1-\epsilon$. We derive our continuum model by using a Taylor series approximation to expand the terms in equation \eqref{eqn: m nonlocal} about $\epsilon = 0$ and extending the domain of validity to $0 \leq a \leq 1$. Dividing through by $\epsilon$ and retaining terms of $O(\epsilon)$ yields the PDE:
\begin{align}
    \frac{\partial m}{\partial t} &= k_L L \Big[ - \frac{\partial}{\partial a} + \frac{\epsilon}{2} \frac{\partial^2}{\partial a^2} \Big] \big[ (1-a) m \big] + k_H H \Big[ \frac{\partial}{\partial a} + \frac{\epsilon}{2} \frac{\partial^2}{\partial a^2} \Big] \big[ a m \big] - (1+\gamma) m. \label{eqn: pde taylor}
\end{align}
Equation \eqref{eqn: pde taylor} is an advection-diffusion equation. In conservation form, it reads:
\begin{align}
    \frac{\partial m}{\partial t} &= \frac{\partial}{\partial a} \Big[ D(a,t) \frac{\partial m}{\partial a} - v(a,t) m \Big] - (1+\gamma) m, \label{eqn: pde conservation}
\end{align}
where the lipid content velocity, $v$, and diffusion coefficient, $D$, are defined as follows:
\begin{align}
    v(a,t) &:= k_L  L (1 - a) - k_H H  a + \frac{\epsilon}{2}(k_L  L - k_H H), \label{eqn: v} \\
    D (a, t) &:= \frac{\epsilon}{2}\big( k_L L (1-a) + k_H H a \big). \label{eqn: D}
\end{align}

Appropriate boundary conditions are determined by considering the dynamics of the dimensionless macrophage population, $M(t)$, which, in the continuum formulation, we identify with the integral:
\begin{align}
    M(t) \sim \int_0^1 m(a,t) \, da. 
\end{align}
Integrating equation \eqref{eqn: pde conservation} with respect to $a \in (0,1)$, we deduce that:
\begin{align}
    \frac{dM}{dt} &= \Big( D \frac{\partial m}{\partial a} - v m \Big) \Big|_{a = 0}^{a = 1} - (1 + \gamma) M. \label{eqn: pde integrated}
\end{align}
To ensure consistency between equations \eqref{eqn: pde integrated} and \eqref{eqn: M nondim}, we set:
\begin{align}
    \Big( D \frac{\partial m}{\partial a} - v m \Big) \Big|_{a = 0} = - \frac{L}{L_c + L}, \qquad \text{and } \qquad \Big( D \frac{\partial m}{\partial a} - v m \Big) \Big|_{a = 1} = 0. \label{eqn: bconds}
\end{align}
The boundary condition at $a = 0$ ensures that the flux of macrophages at $a = 0$ is given by the macrophage recruitment rate, while the no-flux condition at $a = 1$ ensures that macrophages cannot exceed the maximal lipid load $a = 1$. Finally, we impose the initial condition:
\begin{align}
    m(a,0) &= 0, \label{eqn: pde init}
\end{align}
and close the continuum model by coupling the PDE to the ODE subsystem \eqref{eqn: M nondim}-\eqref{eqn: L nondim}. 

% In summary, the full continuum model is given by coupling the ODE subsystem \eqref{eqn: M nondim}-\eqref{eqn: L nondim} to the PDE initial value problem:
% \begin{align}
%     \frac{\partial m}{\partial t} &= \frac{\partial}{\partial a} \Big[ D(a,t) \frac{\partial m}{\partial a} - v(a,t) m \Big] - (1+\gamma) m, & m(a,0) &= 0, \label{eqn: continuum pde}
% \end{align}
% where $v$ and $D$ are defined in equations \eqref{eqn: v} and \eqref{eqn: D} respectively, and imposing the following boundary conditions:
% \begin{align}
%     \Big( D \frac{\partial m}{\partial a} - &v m \Big) \Big|_{a = 0} = - \frac{L}{L_c + L}, \qquad \text{and } \qquad \Big( D \frac{\partial m}{\partial a} - v m \Big) \Big|_{a = 1} = 0. \label{eqn: continuum bconds}
% \end{align}

\subsubsection{Comparison of discrete and continuum models} \label{sec: comparison}

\begin{figure}
    \centering
    \includegraphics[width=1.0\textwidth]{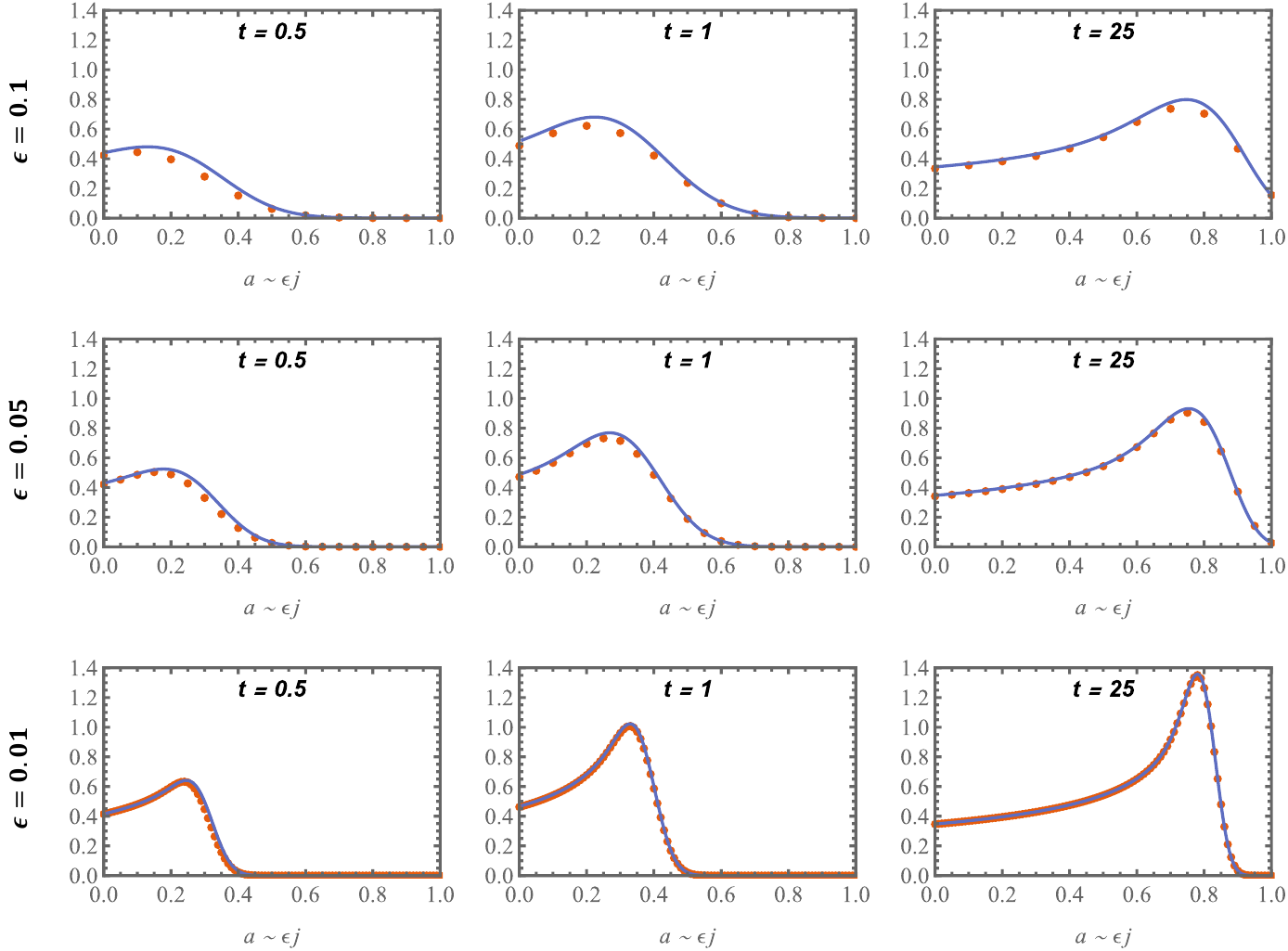}
    \caption{\textbf{Continuum vs. discrete models.} The output for the discrete model \eqref{eqn: m0 nondim}-\eqref{eqn: initconds nondim} is compared to that of the continuum model\eqref{eqn: pde conservation},\eqref{eqn: bconds}, \eqref{eqn: pde init}, \eqref{eqn: M nondim}-\eqref{eqn: initconds nondim} for $\epsilon = 0.1$, $0.05$ and $0.01$ in the top, middle and bottom rows respectively. The distributions are shown at the times $t = 0.5, 1$ and $25$, where the system is approximately at equilibrium. The other parameters are set to: $\sigma_L = 3$, $k_L = 1$, $k_H H = 0.3$, $\kappa = 30$, $\gamma = 0.25$ and $L_c = 1$.}
    \label{fig: continuum vs discrete}
\end{figure}
We generate numerical solutions of the continuum model, defined by equations \eqref{eqn: pde conservation},\eqref{eqn: bconds}, \eqref{eqn: pde init}, \eqref{eqn: M nondim}-\eqref{eqn: initconds nondim}, by using the Method of Lines, with second-order central differencing for $\frac{\partial m}{\partial a}$ and $\frac{\partial^2 m}{\partial a^2}$. The output is compared to that of the discrete model \eqref{eqn: m0 nondim}-\eqref{eqn: initconds nondim} in Figure \ref{fig: continuum vs discrete}. We find that the discrete and continuum models are in excellent agreement, with better consistency for smaller values of $\epsilon$ and late time points, $t$. We note that the value of $\epsilon$ appears to affect the sharpness of the wavefront, but not its location, with smaller values of $\epsilon$ yielding sharper wavefronts. These observations are consistent with equations \eqref{eqn: v} and \eqref{eqn: D}, which show that the diffusion coefficient, $D(a,t)$, is proportional to $\epsilon$, while the velocity, $v(a,t)$, is an $\mathcal{O}(1)$ quantity. 

We quantify the deviation between the discrete and continuum models at time $t$ by defining the mean-squared error (MSE):
\begin{align}
    \text{MSE}(t; \, \epsilon) &:= \frac{1}{J} \sum_{j=0}^{J} \big(J m_j(t) - m(j/J, t) \big)^2, \label{eqn: MSE}
\end{align}
where $J = \epsilon^{-1}$, and $m_j(t)$ and $m(\cdot, t)$ are the discrete and continuum distributions respectively. In Appendix C we plot the MSE across a number of sample times $t$ and values of $\epsilon$. The good agreement between the discrete and continuum models is reflected in that $\text{MSE} < 2 \times 10^{-3}$ for all samples. The largest deviation between the models occurs at early times $0 < t < 1$. This transient difference is suppressed on an $\mathcal{O}(1)$ timescale.

\subsection{Asymptotic analysis at steady state} \label{sec: asymptotics}

Let $m^\star(a) = \lim_{t \rightarrow \infty} m(a,t)$ denote the steady lipid distribution for the continuum model. Setting the time derivative to zero in equation \eqref{eqn: pde conservation} and rearranging, we have that:
\begin{align}
    \frac{\epsilon}{2} \big[ a_c - (2a_c - 1) a \big] \frac{d^2 m^\star}{d a^2} + \big[ \epsilon (2 a_c - 1) + (a - a_c) \big] \frac{d m^\star}{d a} - p m^\star &= 0, \label{eqn: m steady}
\end{align}
where the constants $a_c$ and $p$ are given by:
\begin{align}
    a_c &:= \frac{k_L L^\star}{k_L L^\star+k_H H^\star}, & p &:= \frac{1+\gamma}{k_L L^\star + k_H H^\star} - 1. \label{eqn: ac p defs}
\end{align}
Recall that $L^\star > 0$ and $H^\star > 0$ are the steady state values of $L(t)$ and $H(t)$ respectively. It follows that $0 < a_c < 1$ and $p > -1$. In what follows, we view $a_c$ and $p$ as $\mathcal{O}(1)$ quantities.

At steady state the boundary conditions \eqref{eqn: bconds} are:
\begin{align}
    \frac{\epsilon}{2} \frac{d m^\star}{da} - \Big[ 1 + \frac{\epsilon}{2} \Big( \frac{2a_c - 1}{a_c} \Big) \Big] m^\star &= -R, & \text{at } a = 0,  \label{eqn: bcond1 steady}\\
    \frac{\epsilon}{2} \frac{d m^\star}{da} + \Big[ 1 - \frac{\epsilon}{2} \Big( \frac{2a_c - 1}{1 - a_c} \Big) \Big] m^\star &= 0, & \text{at } a = 1, \label{eqn: bcond2 steady}
\end{align}
where $R := \frac{1}{k_L (L^\star + L_c)}$ is a rescaled recruitment rate. Equations \eqref{eqn: m steady}-\eqref{eqn: bcond2 steady} define a singular perturbation problem with respect to the small parameter $\epsilon \ll 1$. We use the method of matched asymptotics to construct a uniformly valid approximate solution for $m^\star(a)$ in the limit $\epsilon \rightarrow 0$.

By setting $\epsilon = 0$ in equations \eqref{eqn: m steady}, \eqref{eqn: bcond1 steady}, \eqref{eqn: bcond2 steady}, we deduce that the leading order outer solutions satisfy:
\begin{align}
    (a-a_c) \frac{d m^\star}{da} - pm^\star &= 0 \label{eqn: m outer}
\end{align}
and are subject to the Dirichlet conditions:
\begin{align}
    m^\star(0) &= R, & m^\star(1) &= 0. \label{eqn: bconds outer}
\end{align}
The general solution to equation \eqref{eqn: m outer} is given by:
\begin{align}
    m^\star(a) &= K \Big( 1 - \frac{a}{a_c} \Big)^p, \label{eqn: outer sol general}
\end{align}
where $K$ is a constant of integration. Since we cannot enforce both boundary conditions \eqref{eqn: bconds outer} simultaneously, we deduce that there are two outer solutions:
\begin{align}
    m^\star_{\text{left}}(a) &:= R \Big( 1 - \frac{a}{a_c} \Big)^p,  &  m^\star_{\text{right}}(a) &:= 0. \label{eqn: left right outer sols}
\end{align}
The left solution $m^\star_{\text{left}}(a)$ is well-defined for $0 \leq a < a_c$, while $m^\star_{\text{right}}(a)$ holds in a neighbourhood of $a = 1$. We form the following composite solution:
\begin{align}
    m^\star_{\text{outer}}(a) &:= 
    \begin{cases}
        R \Big(1 - \frac{a}{a_c} \Big)^p, & 0 \leq a < a_c, \\
        0, & a_c < a \leq 1,
    \end{cases} \label{eqn: outer composite sol}
\end{align}
which ensures continuity in $m^\star_{\text{outer}}$ at $a = a_c$ for $p > 0$.

\begin{figure}
    \centering
    \includegraphics[width=1.0\textwidth, height=0.2\textheight]{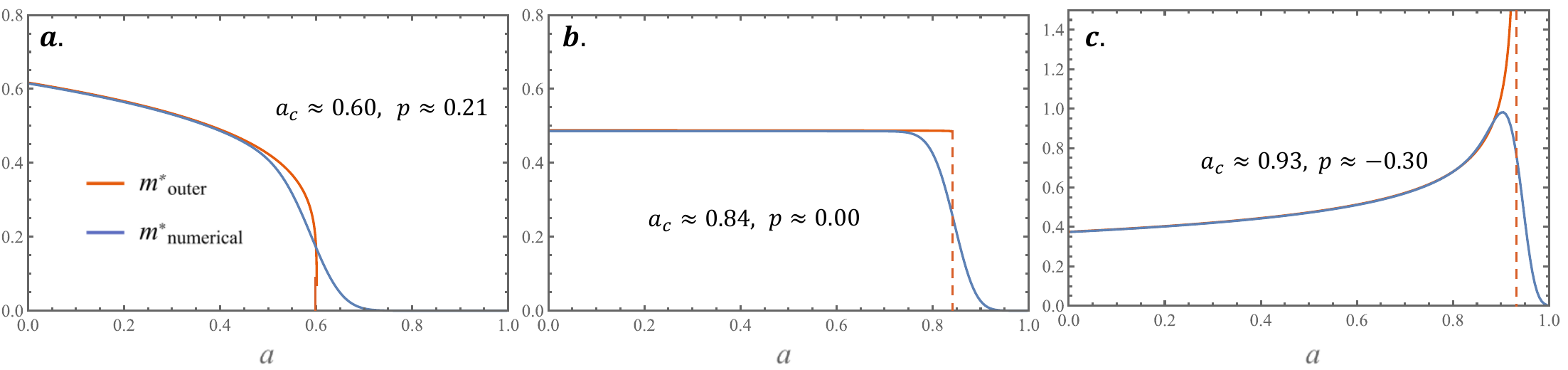}
    \caption{\textbf{Composite outer solution.} The outer solution \eqref{eqn: outer composite sol} is compared with the numerical solution for $m^\star(a)$ for three values of the LDL influx, $\sigma_L$. The numerical solution is generated by solving the time-dependent system \eqref{eqn: pde conservation},\eqref{eqn: bconds}, \eqref{eqn: pde init}, \eqref{eqn: M nondim}-\eqref{eqn: initconds nondim} to $t = 200$. We use $\sigma_L = 1.3$, $1.845$ and $2.5$ in cases a,b and c respectively. The red dashed line in cases b and c indicate the discontinuity of $m^\star_{\text{outer}}$ at $a = a_c$. We fix $\sigma_H = 1$ and set the remaining parameters to the default values stated in equations \eqref{eqn: parvals}.}
    \label{fig:outer sol}
\end{figure}

Numerical solutions of the continuum model indicate that $m^\star_{\text{outer}}(a)$ is a good approximation to $m^\star(a)$ away from the critical value $a = a_c$ (see Figure \ref{fig:outer sol}). For $p > 0$, as in Fig.\ref{fig:outer sol}a, the outer solution \eqref{eqn: outer composite sol} is continuous at $a = a_c$, but the numerical solution has a smoother transition towards the zero solution as $a$ increases. For $p = 0$, shown in Fig.\ref{fig:outer sol}b, the solution \eqref{eqn: outer composite sol} is piecewise constant with a discontinuity at $a = a_c$. By contrast, the numerical solution is continuous for all $0 < a < 1$ and takes on a sigmoid profile. For $p < 0$, as in Fig.\ref{fig:outer sol}, the solution \eqref{eqn: outer composite sol} diverges as $a \rightarrow a_c^{-}$, while the numerical solution has a local maximum near $a = a_c$ in this case.  

These observations suggest that we may use $p$ and $a_c$ to characterise the solution $m^\star(a)$ of the full equation \eqref{eqn: m steady}. The value of $p$ determines the geometric features of $m^\star(a)$. For $p > 0$, the solution is a decreasing function of $a$. The curvature is determined by whether $p > 1$, giving a concave-up $m^\star(a)$ for $0 < a \lesssim a_c$, or $0 < p < 1$, where $m^\star(a)$ is concave-down for $0 < a \lesssim a_c$. If $p = 0$, then $m^\star(a)$ is a sigmoid and constant away from $a = a_c$. Finally, if $p < 0$, then $m^\star(a)$ attains a local maximum near $a = a_c$, with the sharpness of the peak growing as $p \rightarrow (-1)^{+}$. 

\begin{figure}
    \centering
    \includegraphics[width=\textwidth]{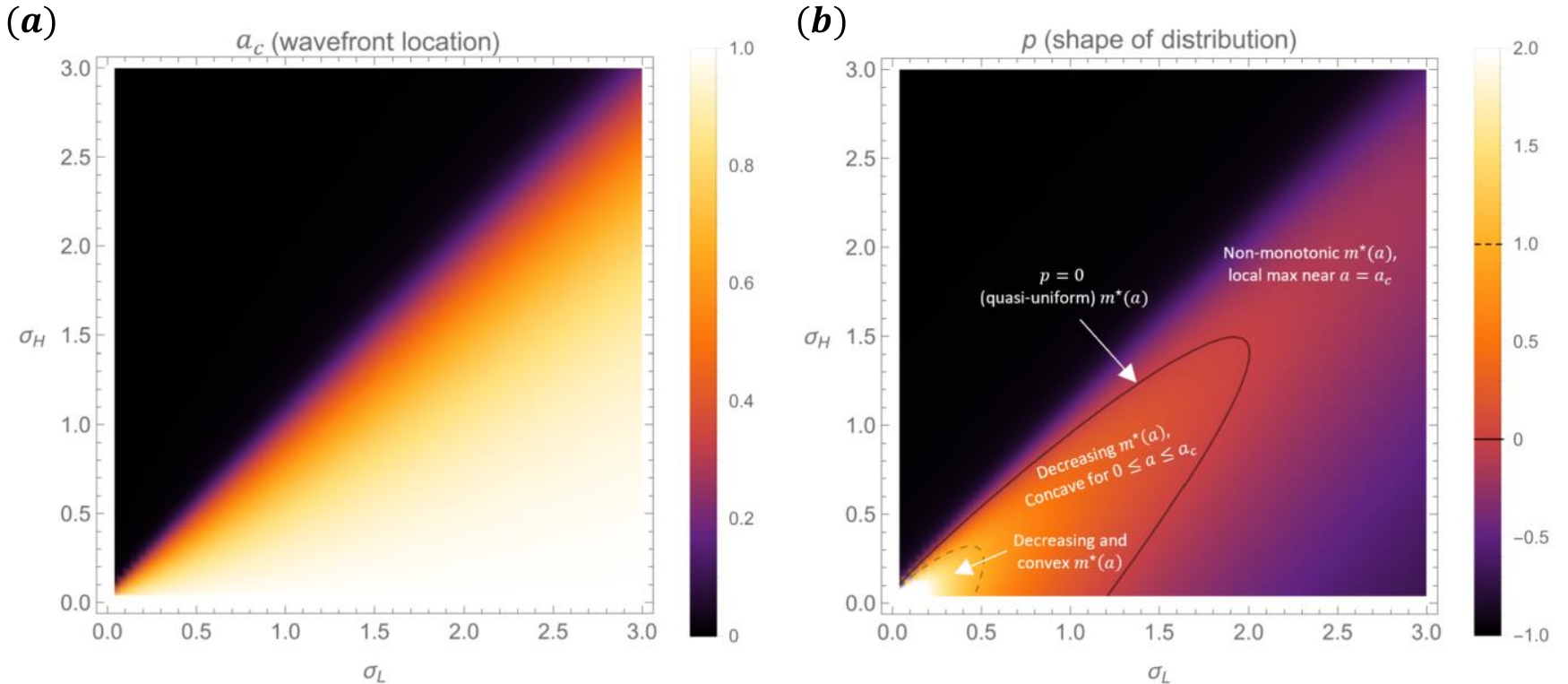}
    \caption{\textbf{Characterisation of the equilibrium macrophage lipid distribution, $m^\star(a)$, in terms of LDL influx, $\sigma_L$, and HDL influx, $\sigma_H$.} Plot (a) shows the value of $a_c$, which approximates the lipid content of the most lipid-laden macrophage at equilibrium. Plot (b) indicates the value of the parameter $p$, which determines the shape of the distribution as indicated. Parameter values are set according to equations \eqref{eqn: parvals}.}
    \label{fig:m(a) characterisation}
\end{figure}

We show how $p$ and $a_c$ depend on $\sigma_L$ and $\sigma_H$ in Figure \ref{fig:m(a) characterisation}. We find that the contours of constant $a_c$ (Fig.\ref{fig:m(a) characterisation}a) are approximately linear in $(\sigma_L, \sigma_H)$ space, and indicate that $a_c$ increases sensitively with $(\sigma_L - \sigma_H)$. By contrast, the value of $p$ (Fig.\ref{fig:m(a) characterisation}b) is non-monotonic in both $\sigma_L$ and $\sigma_H$, and diverges at the origin: $p \rightarrow \infty$ as $(\sigma_L, \sigma_H) \rightarrow (0,0)$. The critical contours, $p = 0$ and $p = 1$, are indicated so that we may partition $(\sigma_L, \sigma_H)$ space according to the qualitative behaviour of the steady state profile of $m^\star(a)$. Consistent with the numerical results of Section \ref{sec: dynamics}, we find that $m^\star(a)$ has a peaked profile ($p < 0$) when either $\sigma_L$ or $\sigma_H$ are large enough. Comparison with the corresponding value of $a_c$ reveals that peaked profiles can occur with the local maximum at low or high lipid loads (location of peak is approximately $a_c$). Furthermore, $m^\star(a)$ only adopts a decreasing and convex profile when $\sigma_L$ and $\sigma_H$ are both sufficiently small.

To attain a uniformly valid approximation for $m^\star(a)$ over $0 < a < 1$, we define an interior variable $a = a_c + \epsilon^n \sqrt{2 a_c(1-a_c)} \, \Tilde{a}$ where $n > 0$ is to be specified and the coefficient of $\Tilde{a}$ is chosen to simplify the interior equation. Under this change of variables, equation \eqref{eqn: m steady} becomes:
\begin{align}
    \frac{\epsilon^{1-2n}}{2} \Big[ 1 + \epsilon^n  \frac{1-2a_c}{2a_c(1-a_c)}  \Tilde{a} \Big] \frac{d^2 m^\star}{d \Tilde{a}^2} + \Big[ \epsilon^{1-n} \frac{2a_c-1}{\sqrt{2a_c(1-a_c)}} + \Tilde{a} \Big] \frac{d m^\star}{d \Tilde{a}} - p m^\star &= 0,
\end{align}
for $- \infty < \Tilde{a} < \infty$. In order to match to the two outer solutions we choose $n$ such that a term proportional to $\frac{d^2 m^\star}{d \Tilde{a}^2}$ becomes comparable to the $O(1)$ terms. The smallest value of $n$ for which this occurs is $n = \frac{1}{2}$, which gives the following equation at leading order:
\begin{align}
    \frac{1}{2} \frac{d^2 m^\star}{d \Tilde{a}^2} + \Tilde{a} \frac{d m^\star}{d \Tilde{a}} - p m^\star &= 0. \label{eqn: interior eqn}
\end{align}
The general solution to equation \eqref{eqn: interior eqn} is:
\begin{align}
\begin{split}
    m^\star_{\text{interior}}(\Tilde{a}) &= e^{-\Tilde{a}^2} \Big[ K_1 H_{-1-p}(\Tilde{a})  + K_2 \,  {}_1F_1 \Big( \frac{1+p}{2}, \frac{1}{2}, \Tilde{a}^2 \Big) \Big], \label{eqn: interior sol gen}
\end{split}
\end{align}
where $H$ and ${}_1F_1$ are the Hermite and Kummer confluent hypergeometric functions respectively (\cite{1F1, HermiteH}). The constants of integration, $K_1$ and $K_2$, are determined by matching $m^\star_{\text{interior}}$ to the outer solutions $m^\star_{\text{left}}$ and $m^\star_{\text{right}}$ respectively. We note that equation \eqref{eqn: interior sol gen} satisfies the following asymptotic relation as $\Tilde{a} \rightarrow \infty$:
\begin{align}
    m^\star_{\text{interior}} &\sim K_2 \frac{\pi^{1/2}}{\Gamma((1+p)/2)} \Tilde{a}^p \nonumber \\
    &= K_2 \frac{\pi^{1/2}}{\Gamma((1+p)/2)} \epsilon^{-p/2} (a-a_c)^p.
\end{align}
Comparing with $m^\star_{\text{right}}(a)$ shows that we must take $K_2 = 0$. As $\Tilde{a} \rightarrow - \infty$, we use that:
\begin{align}
    m^\star_{\text{interior}} &\sim -K_1 \pi^{1/2} \Gamma(-p) \sin(p \pi) (-\Tilde{a})^p \nonumber \\
    &= -K_1 \pi^{1/2} \Gamma(-p) \sin(p \pi) a_c^p \epsilon^{-p/2} \Big( 1 - \frac{a}{a_c} \Big)^p,
\end{align}
and compare with $m^\star_{\text{left}}(a)$ to deduce that the matched interior solution is:
\begin{align}
    m^\star_{\text{interior}}(\Tilde{a}) &= K_1 e^{-\Tilde{a}^2} H_{-1-p}(\Tilde{a}), \qquad \text{where } \quad K_1 = \frac{- R \sqrt{\pi} [2\epsilon a_c (1-a_c )]^{p/2}}{ a_c^p \Gamma (-p) \sin (p \pi)}. \label{eqn: m interior sol}
\end{align}
We note that $\Gamma (-p) \sin (p \pi) < 0$ for all $p > -1$ and so $K_1 > 0$.

\begin{figure}
    \centering
    \includegraphics[width=0.6\textwidth]{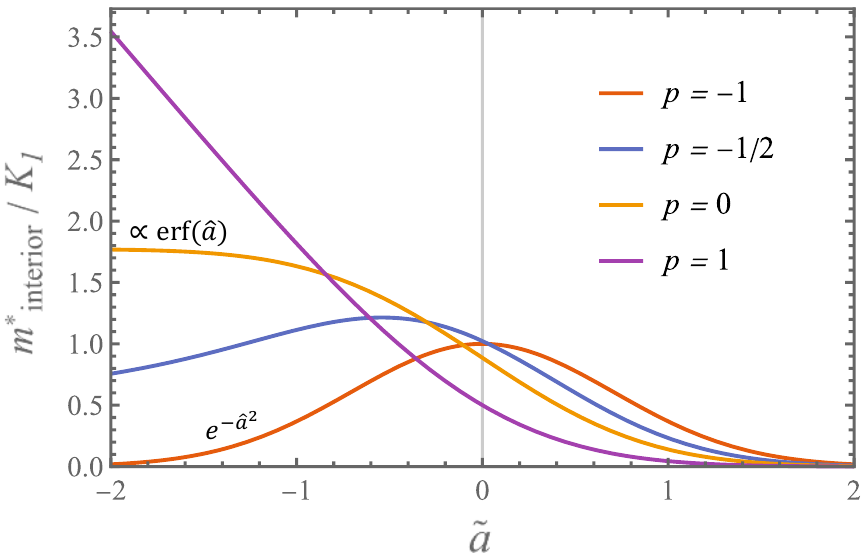}
    \caption{\textbf{The interior solution}. The function $e^{-\tilde{a}^2} H_{-1-p}(\tilde{a})$ from the interior solution \eqref{eqn: m interior sol} is shown for the values $p = -1$ (limiting case, Gaussian), $-1/2$, $0$ (sigmoid) and $1$.}
    \label{fig:interior plots}
\end{figure}

The function $m^\star_{\text{interior}}/K_1$ from \eqref{eqn: m interior sol} is plotted for various values of $p$ in Figure \ref{fig:interior plots}. We divide by $K_1$ since it affects only the vertical scale of the interior solution. The function becomes increasingly asymmetric as $p$ increases, transitioning from a bell curve at $p = -1$, to a sigmoid at $p = 0$, and eventually a smoothed corner-like shape for $p > 0$. Consistent with these observations, we find that $m^\star_{\text{interior}}$ reduces exactly to a Gaussian in the limit $p \rightarrow (-1)^{+}$ and a sigmoid at $p = 0$:
\begin{align}
        m^\star_{\text{interior}}(\Tilde{a}) &\rightarrow K_1 e^{-a^2}, & p &\rightarrow (-1)^{+}; \\
        m^\star_{\text{interior}}(\Tilde{a}) &= \frac{K_1 \sqrt{\pi}}{2} \text{erf}(\Tilde{a}), & p &= 0.
\end{align}

\begin{figure}
    \centering
    \includegraphics[width = 0.6\textwidth]{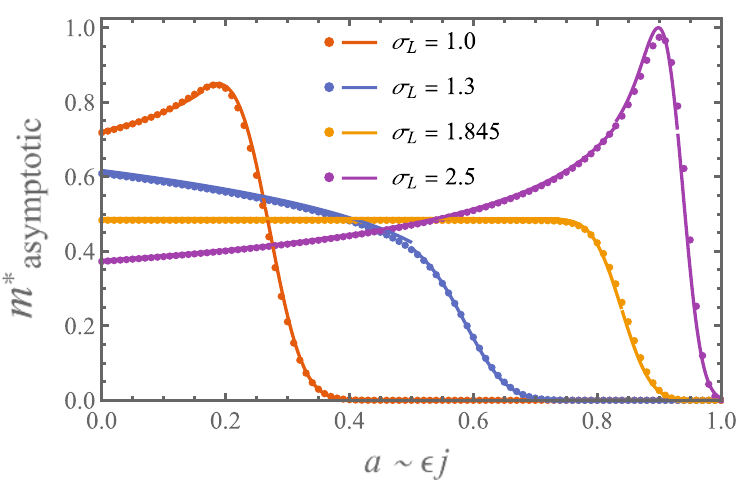}
    \caption{\textbf{Asymptotic solution vs discrete model output.} The asymptotic solution of the continuum model, defined by equation \eqref{eqn: m asymptotic sol}, is plotted with the discrete model steady state distribution, $J m_j^\star$. We use $\sigma_H = 1$ and set the remaining parameter values according to equations \eqref{eqn: parvals}.}
    \label{fig:asymp vs discrete}
\end{figure}

A leading order approximation for $m^\star(a)$ that is uniformly valid over $0 \leq a \leq 1$ is obtained by substituting $\Tilde{a} = [2\epsilon a_c(1-a_c)]^{-1/2}(a-a_c)$ into equation \eqref{eqn: m interior sol}:
\begin{align}
    m^\star_{\text{asymptotic}}(a) = K_1 \exp\bigg[ - \frac{(a-a_c)^2}{2 \epsilon a_c (1-a_c)} \bigg] H_{-1-p} \bigg( \frac{a-a_c}{\sqrt{2 \epsilon a_c(1-a_c)}} \bigg). \label{eqn: m asymptotic sol}
\end{align}
We compare the matched solution \eqref{eqn: m asymptotic sol} to the discrete model output in Figure \ref{fig:asymp vs discrete}. The plots show an excellent agreement between the asymptotic solution and discrete model.

\section{Discussion} \label{sec: Discussion}

% What have we done? What is our model, and what makes it unique?
% you have now presented the model - think carefully about tenses, remembering that all the results and analysis have now been presented. You can use present tense when talking about the model. 
In this paper we presented a %new 
discretely structured population model for lipid accumulation in atherosclerotic plaque macrophages. The model is formulated in terms of the amount of extracellular lipid, $L(t)$, the lipid capacity of the HDL particles, $H(t)$, and the macrophage population, $m_j(t)$, which is structured by an index, $j$, according to lipid content. A key feature of the model is that lipid uptake and offloading events cause a finite change, $\Delta a$, to macrophage lipid content. Our discretely-structured model formulation contrasts to the continuum approach proposed by %in the lipid-structured model of 
Ford et al. \cite{ford2019lipid}. It can also be reduced to a simpler subsystem of ODEs for $L(t)$, $H(t)$, the total number of macrophages, $M(t)$, and the total macrophage lipid content, $A_M(t)$. 

A defining feature of our model is the treatment of macrophage lipid uptake and offloading: the uptake rate is assumed to decrease linearly with lipid content, $k_L (1 - j/J)$, 
while the offloading rate increases linearly,
according to $k_H j/J$. 
These assumptions are consistent with current understanding that macrophages have a finite capacity for phagocytosis due to intracellular negative-feedback signalling \cite{zent2017maxed}, and that intracellular cholesterol accumulation promotes the up-regulation of transporters that mediate cholesterol efflux \cite{remmerie2018macrophages}. An important consequence of our model formulation is that macrophages have a maximal lipid content, $j \leq J$, since the uptake rate decreases to zero at $j = J$. By contrast, the model proposed by Ford et al. \cite{ford2019lipid}) assumes that lipid uptake and offloading are independent of lipid content, and that there is no upper bound on macrophage lipid content. We further note that our simplifying assumption of linear dependence in $j$ for the uptake and offloading rates ensures the existence of a closed ODE subsystem: \eqref{eqn: L nondim}-\eqref{eqn: initconds nondim}. If the lipid-dependent rates were  nonlinear in $j$, the dynamics of $A_M(t)$ would not be expressible in terms of $M(t)$, $A_M(t)$, $L(t)$ and $H(t)$, and opportunities for mathematical analysis would be limited.

Our model analysis focused on the constants $\sigma_L$ and $\sigma_H$ which determine the influx of LDL lipid and HDL capacity respectively. These parameters are expected to correlate strongly with blood LDL and HDL particle concentrations. %To achieve the aims of the study outlined in Section \ref{sec: intro}, w
We used the ODE subsystem to determine how changing the values of $\sigma_L$ and $\sigma_H$ impacts overall plaque composition and macrophage lipid distribution. %by analysing the functions $m_j$ and their corresponding continuum formulation $m(a,t)$.

Numerical solutions showed that, at long times, the plaque develops to a nonzero equilibrium. The dynamics proceed through various stages that are detailed in Section \ref{sec: dynamics}. Considering firstly the ODE subsystem, we note that $M(t)$ increases with $t$, but $A_M(t)$, $L(t)$ and $H(t)$ may be non-monotonic; the plaque has a steadily increasing population of macrophages, but the dynamics of its intracellular lipid, extracellular lipid and unloaded HDL content may not be monotonic. Further, the subsystem dynamics are sensitive to $(\sigma_L - \sigma_H)$, the influx of LDL lipid relative to HDL capacity. As expected, the higher the entry rate of LDL lipid to HDL capacity, the higher the lipid accumulation in the plaque. The macrophage lipid distribution $m_j(t)$ evolves in a wave-like manner towards its steady state solution. %In accordance with the non-monotonic subsystem dynamics, t
The wavefront velocity changes over time and may change sign (i.e., the front travels back to lower lipid loads) if $H(t)$ is sufficiently large; average macrophage lipid loads may decrease when offloading to HDL exceeds phagocytosis. We note that the equilibrium lipid distribution, $m_j$, takes one of several qualitatively distinct shapes. These profiles may be monotone decreasing (concave or convex), quasi-uniform and or uni-modal. The equilibrium profile is sensitive to the individual values of $\sigma_L$ and $\sigma_H$, rather than their difference, $(\sigma_L - \sigma_H)$. 

Steady state analysis of the subsystem \eqref{eqn: M nondim}-\eqref{eqn: H nondim} shows that the model has a unique physically-realistic equilibrium. Analysis of the steady state solutions (see Figure \ref{fig:steady contours}) indicated that the number of macrophages, $M^\star$, amounts of intracellular lipid, $A_M^\star$, extracellular lipid, $L^\star$, and HDL capacity, $H^\star$, depend sensitively on the difference $(\sigma_L - \sigma_H)$. Hence, the model predicts that plaque composition is strongly dependent on the difference between the influx of LDL lipid relative to HDL capacity. More specifically, the plots suggest that $(\sigma_L, \sigma_H)$ space can be partitioned into three regions. Regions in which $\sigma_L - \sigma_H \lesssim 0$ correspond to healthy tissues in which the HDL capacity greatly exceeds the amounts of ingested and extracellular lipid.
Where $0 \lesssim \sigma_L -\sigma_H \lesssim 5$, the HDL capacity is typically exhausted and plaques contain intermediate quantities of macrophages and lipid. Finally, regions in which $\sigma_L - \sigma_H \gtrsim 5$ correspond to unhealthy plaques that are dominated by extracellular lipid and macrophages with lipid loads near capacity. 

% Indeed, the asymptotic solution $L^\star_{2.+}$ (equations \eqref{eqn: L2 asymp} and \eqref{eqn: L2 coeffs}) in the limit $\delta_L, \delta_H \ll 1$ (negligible non-MDM loss of extracellular lipid and HDL) only depends on $\sigma_L$ and $\sigma_H$ through the difference $\sigma_L - \sigma_H$ and weighted difference $(1+\gamma)\sigma_L - \sigma_H$. 
%
% \begin{itemize}
% \item this is quite technical - I would describe in words and include math in brackets (I think)
% \end{itemize}
% %
% The plots further suggest that $(\sigma_L, \sigma_H)$ space can be partitioned into three regions. Regions in which $\sigma_L - \sigma_H \lesssim 0$ correspond to healthy tissues which are dominated by HDL rather than macrophages or lipid. 
% \begin{itemize}
% \item what does a tissue which is dominated by HDL look like? I don't understand
% \end{itemize}
%
% Where $0 \lesssim \sigma_L -\sigma_H \lesssim 5$, the HDL capacity is typically exhausted and plaques contain intermediate quantities of macrophages and lipid. Finally, regions in which $\sigma_L - \sigma_H \gtrsim 5$ correspond to unhealthy plaques that are dominated by extracellular lipid and macrophages with lipid loads near capacity. 

Equations \eqref{eqn: m0 nondim}-\eqref{eqn: mJ nondim} for the dynamics of $m_j(t)$, $j = 0, 1, \dots, J$, are difficult to analyse directly. Nonetheless, we show in Appendix B that an exact solution \eqref{eqn: m_j formula} for the equilibrium values, $m_j^\star$, can be constructed via the method of generating functions. Unfortunately, the solution is too complicated to extract any biological insight. Instead, we analysed the macrophage lipid distribution by deriving a continuum analogue of the discrete model. The continuum model casts the macrophage population as a single function, $m(a,t)$, where the continuous independent variable $0 \leq a  \leq 1$ represents lipid content. The distribution $m(a,t)$ is governed by an advection-diffusion PDE \eqref{eqn: pde conservation} with appropriate boundary conditions \eqref{eqn: bconds}.

% Instead, we analysed the macrophage lipid distribution by recasting the discrete populations $m_j(t)$, $j = 0, 1, \dots, J$, as a single function $m(a,t)$ that is structured by a real variable $0 \leq a \leq 1$, representing lipid content. 

% \begin{itemize}
% \item I think there is too much detail here - just say that you derive continuum analogue of discrete model
% \end{itemize}
% The discrete equations \eqref{eqn: m0 nondim}-\eqref{eqn: mJ nondim} are recast, using a second-order Taylor series expansion in the limit $\epsilon = J^{-1} \rightarrow 0$, as an advection-diffusion PDE \eqref{eqn: pde conservation} for the dynamics of $m(a,t)$, with the boundary conditions \eqref{eqn: bconds} at $a = 0, 1$. The results shown in Figure \ref{fig: continuum vs discrete} indicate that the continuum approximation is in good agreement with the discrete model output for $\epsilon < 10^{-1}$, with better consistency at larger lipid contents $a$, and time $t$. 

We used the method of matched asymptotics to construct a leading order approximation to the equilibrium macrophage lipid distribution, $m^\star(a)$, in the limit $\epsilon = J^{-1} \rightarrow 0$. %In addition to $\epsilon > 0$,% 
We found that $m^\star(a)$ depends on two parameter groupings: $0 < a_c < 1$ and $p > -1$, defined in \eqref{eqn: ac p defs}. The constant $a_c$ approximates the maximal lipid load attained by an individual macrophage at equilibrium, and represents the macrophage lipid load where uptake balances offloading. Near $a = a_c$, the advection velocity is $\mathcal{O}(\epsilon)$, and comparable to the diffusion terms. The quantity $p$ determines the shape of $m^\star(a)$. Using the outer solution \eqref{eqn: outer composite sol} as a guide, we found that $m^\star(a)$ is decreasing and convex for $p > 1$, decreasing and concave for $0 < a \lesssim a_c$ when $0 < p < 1$, quasi-uniform for $p = 0$, and unimodal with a local maximum near $a = a_c$ for $p < 0$. We note that the outer solution is discontinuous at $a = a_c$ for $p \leq 0$, and in the first derivative for $0 < p < 1$ (see Figure \ref{fig:outer sol}). Analysis within an interior region of $O(\epsilon^{1/2})$ width at $a = a_c$ is needed to construct a uniformly valid approximation for $m^\star(a)$, resulting in the solution \eqref{eqn: m asymptotic sol}. These observations highlight the importance of including second-order terms in the continuum approximation to capture the qualitative behaviour of the underlying discrete model. The second-order terms provide a diffusive effect that smooths the outer solution in the interior region where the advection velocity vanishes at first order.

Figure \ref{fig:m(a) characterisation} shows that the parameter groupings $a_c$ and $p$ are sensitive to the influxes of LDL lipid $(\sigma_L)$ and HDL capacity $(\sigma_H)$. Like the steady state values of the reduced subsystem variables, the constant $a_c$ increases with $(\sigma_L - \sigma_H)$; the higher the influx of LDL lipid relative to HDL capacity, the higher the lipid load of the most lipid-laden macrophage in the population. In particular, by comparing Figures \ref{fig:summary steady} and \ref{fig:m(a) characterisation} we see that $a_c$ and $\mu^\star$ (average macrophage lipid burden, or equivalently the fractional defect in phagocytosis due to lipid loading) are highly correlated. Noting that $a_c \gtrsim 0.99$ and $\mu^\star \gtrsim 60\%$ for $\sigma_L - \sigma_H \gtrsim 3$, the model indicates that plaque macrophages can expect at least a $60\%$ decrease in their population-averaged phagocytosis rate (purely due to lipid accumulation) if macrophages with lipid content near the capacity value, $a = 1$, are detected. These results show that
%, since macrophages have a finite capacity, 
the phagocytic efficiency of macrophage populations \textit{in vivo} may be markedly reduced by intracellular lipid accumulation. In particular, intracellular lipid accumulation may serve as a partial explanation for the defective efferocytosis often observed in atherosclerotic plaques \cite{yurdagul2018mechanisms, ge2022efferocytosis}. The model also indicates that this hypothesis may be verified experimentally if macrophages with lipid content near capacity are detected \textit{in vivo}.

By contrast, the parameter $p$ is non-monotonic in both $\sigma_L$ and $\sigma_H$, and, hence, the shape of the distribution $m^\star(a)$ depends on the individual influxes of LDL lipid and HDL capacity. In particular, $m^\star(a)$ may adopt a peaked or quasi-uniform profile at low or high lipid loads. These results suggest that, unlike the maximally attained lipid load $a_c$, the geometric features of the macrophage lipid distribution, such as the concavity and existence of local maximum, may be poor indicators of plaque progression.

There are many ways in which our model could be extended. For instance the model assumes that the rates of macrophage death and emigration are constant. However, experimental observations indicate that lipid acquisition can induce cytotoxic macrophage death \cite{yin2021cellular}, and lipid-laden macrophages are less capable of emigration \cite{chen2019lysophosphatidic}. These effects could be accommodated by viewing $\beta$ as an increasing function of $j$ and $\gamma$ as a decreasing function of $j$. Lipid-dependent death and emigration rates are the focus of a study by Watson et al. (\cite{watson2022lipid}). Based upon their results, we expect lipid-dependent death and emigration rates to increase the overall plaque lipid burden relative to the prediction of our model, with more macrophages bearing high lipid loads. Our model further assumes that macrophage proliferation is negligible. However, experiments show that, although proliferation is more significant in the later stages of plaque development, it is still detectable in early plaques (Robbins et al. \cite{robbins2013local}). The effect of proliferation on lipid accumulation in atherosclerosis is the focus of the model by Chambers et al. (\cite{chambers2022lipid}). Based upon their results, we anticipate that incorporating proliferation into our model will decrease the average lipid burden of the macrophages, opposing the effects of lipid-dependent cell death and emigration. 

We could also extend the model to account for spatial variation in the dependent variables. This would allow us to study the spatial distribution of intracellular lipid, and to determine how the location and shape of the extracellular lipid core change as model parameters vary. Such an extended model would be similar in form to the spatially resolved phenotype-structured models of tumour growth by Celora et al. \cite{celora2023spatio} and Fiandaca et al. \cite{fiandaca2022phenotype}. Based on these works, we anticipate that analysis of a spatially explicit lipid-structured model would rely primarily on numerical simulations.

% A spatially-resolved model also provides a more natural framework in which to incorporate the model kinetics. 
% \begin{itemize}
% \item why does a spatially-resolved model provide more natural framework within which to incorporate model kinetics? what kinetics are envisaged here?
% \item also please cite relevant math models which account for spatial and phenotypic variation eg Guilia's work
% \item in next sentences I think you are describing the sorts of additional (boundary) conditions that would need to be imposed if model extended to include spatial variation - ie these aren't just an optional add-on to the current model
% \end{itemize}
% The influx of LDL, HDL and macrophages from the bloodstream could be modelled by endothelial boundary conditions rather than via internal source terms. Similarly, macrophage emigration could be incorporated as a boundary condition so that only macrophages near the plaque boundary can leave the system.

The current model further assumes that the rate of macrophage recruitment increases with the amount of extracellular lipid, $L(t)$. A more realistic treatment of recruitment would account for the production of pro-inflammatory and anti-inflammatory/pro-mediating signals by macrophages, which ultimately requires a treatment of macrophage phenotype. One possibility, explored in Ford et al. \cite{ford2019lipid}, is to assume recruitment increases with macrophage lipid burden. However, this assumption is at odds with experimental studies that indicate lipid-laden macrophages express fewer inflammatory genes compared with macrophages with a lesser lipid burden \cite{kim2018transcriptome}. Another approach, which we will pursue in a future study, is to treat phenotype as a separate structure variable that determines the recruitment rate, and distinguish between pro-inflammatory (e.g. uptake of modified LDL) and anti-inflammatory (e.g. efferocytosis) phagocytosis events. The inclusion of anti-inflammatory effects that dampen macrophage recruitment is likely to complicate the time-dependent behaviour of the macrophage population, $M(t)$, relative to the monotone increase with $t$ observed in our model.

\section{Conclusions} \label{sec: conclusions}

% \begin{itemize}
% \item why a separate section for conclusions? is it really needed?
% \end{itemize}

In this paper we have presented a new lipid-structured model for atherosclerotic plaque macrophages. We showed that the overall plaque composition and the distribution of lipid within the macrophage population are highly sensitive to the influxes of LDL lipid ($\sigma_L$) and HDL capacity $(\sigma_H)$. %, which reflect their respective bloodstream particle concentrations, sensitively affect both . Specifically, we found that plaque lipid content is highly sensitive to the relative influx of LDL to HDL. 
Our results support the notion that plaque lipid content is best reduced by simultaneously lowering the bloodstream concentration of LDL and raising that of HDL. We show further how the qualitative profile of the equilibrium macrophage lipid distribution changes as $\sigma_L$ and $\sigma_H$ vary. These profiles may be monotone decreasing (concave or convex), quasi-uniform or unimodal. Our results also highlight that, since macrophages have a finite capacity, the efficiency of lipid uptake by macrophages may be severely impaired by lipid accumulation. Lipid accumulation in plaque macrophages may therefore serve as a partial explanation for the defective efferocytosis often observed in atherosclerotic plaques, especially if macrophages with lipid content near capacity are observed \textit{in vivo}.

\section*{Declaration of competing interest}

The authors declare that they have no known competing financial interests or personal relationships that could have appeared to influence the work reported in this paper.

\section*{Acknowledgements}

The authors wish to thank Michael G Watson, Laura Chaffey, Gareth Purvis and David R Greaves for helpful conversations that contributed to the ideas in this article. KC acknowledges support from the Oxford Australia Scholarships Fund and Clarendon Scholars' Association. 

% \bibliography{refs}        %use a bibtex bibliography file refs.bib
% \bibliographystyle{unsrt}  %use the plain bibliography style
\newpage
\appendix

\section{Complete statement of polynomial \eqref{eqn: Lstar quintic} for $L^\star$}

Setting the time derivatives to zero in equations \eqref{eqn: M nondim}-\eqref{eqn: H nondim} reveals that $L^\star$ satisfies a quintic polynomial:
\begin{align}
   c_5 (L^\star)^5 +  c_4 (L^\star)^4+  c_3 (L^\star)^3 +  c_2 (L^\star)^2 +  c_1 L^\star + c_0 = 0, \label{eqn: L quintic appendix}
\end{align}
with the following coefficients:
\begin{align}
    c_0 &:= (\gamma +1)^2 L_c^2 \sigma _L (\gamma  \delta _H+\delta _H+k_H (\sigma _H-(\gamma +1) \sigma _L)), \\
    c_1 &:= (\gamma +1) L_c \Big[(\gamma +1) (\gamma +2) L_c \delta _H k_L \sigma _L \nonumber \\
    &\quad + L_c k_H \Big(k_L \sigma _L (\sigma _H-\sigma _L)-\sigma _H \delta _L +2 (\gamma +1) \delta _L \sigma _L\Big) -(\gamma +1) L_c \delta _H \delta _L+\delta _H \\
    &\quad -2 k_H \sigma _L (-\sigma _H+\gamma  \sigma _L+\sigma _L+1)+2 (\gamma +1) \delta _H \sigma _L)+k_H \sigma _H\Big],  \nonumber \\  
    c_2 &:= (\gamma +1) (k_H (-L_c \sigma _H k_L ((\gamma +1) L_c \delta _L+\kappa -2) \nonumber\\
    &\quad +L_c k_L \sigma _L (2 \gamma  (L_c \delta _L+\sigma _H+\kappa -1)+2 L_c \delta _L+2 \sigma _H+\kappa -3)  -2 (\gamma +1) L_c \sigma _H \delta _L \nonumber \\
    &\quad+(\gamma +1) \sigma _L (4 (\gamma +1) L_c \delta _L+\sigma _H-2)-(\gamma +1) \sigma _L^2 (2 L_c k_L+\gamma +1) \nonumber \\
    &\quad -((\gamma +1) L_c \delta _L-1){}^2+\sigma _H)+L_c \delta _H (-(\gamma +1) \delta _L ((\gamma +2) L_c k_L+2 (\gamma +1))-\gamma  \kappa  k_L) \nonumber \\
    &\quad +(\gamma +1) \delta _H \sigma _L (L_c k_L (L_c k_L+2 \gamma +4)+\gamma +1)+(\gamma +1) \delta _H (2 L_c k_L+1)), \\
    c_3 &:= (\gamma +1) \delta _H (-(\gamma +1) (\delta _L (L_c k_L (L_c k_L+2 \gamma +4)+\gamma +1) \nonumber \\
    &\quad -k_L (L_c k_L+2))-\gamma  \kappa  k_L (L_c k_L+1)+(\gamma +1) k_L \sigma _L (2 L_c k_L+\gamma +2)) \nonumber \\
    &\quad +k_H (k_L ((\gamma +1) \sigma _L (\gamma  (4 L_c \delta _L+\sigma _H+2 \kappa -2)+4 L_c \delta _L+\sigma _H+\kappa -3) \nonumber \\
    &\quad -(\gamma +1) \sigma _H (2 (\gamma +1) L_c \delta _L+\kappa -2)-((\gamma +1) L_c \delta _L-1) (2 \gamma  \kappa +(\gamma +1) (L_c \delta _L-2)+\kappa ) \nonumber \\
    &\quad -(\gamma +1)^2 \sigma _L^2)-(\gamma +1) (\kappa -1) L_c k_L^2 (\sigma _H-\sigma _L)+(\gamma +1)^2 (-\delta _L) (2 (\gamma +1) L_c \delta _L \nonumber \\
    &\quad +\sigma _H-2 (\gamma +1) \sigma _L-2)), \\
    c_4 &:= (\gamma +1) \delta _H k_L (k_L (\gamma  (-2 L_c \delta _L-\kappa +\sigma _L+1)-2 L_c \delta _L+\sigma _L+1)-(\gamma +1) (\gamma +2) \delta _L) \nonumber \\
    &\quad -k_H ((\gamma +1) k_L \delta _L (\gamma  (2 L_c \delta _L+\sigma _H+2 \kappa -2 \sigma _L-2)+2 L_c \delta _L+\sigma _H+\kappa -2 \sigma _L-3) \nonumber \\
    &\quad +(\kappa -1) k_L^2 (\gamma  (L_c \delta _L+\sigma _H+\kappa -\sigma _L-1)+L_c \delta _L+\sigma _H-\sigma _L-1)+(\gamma +1)^3 \delta _L^2), \\
    c_5 &:= -[(\gamma +1) k_L \delta _L ((\gamma +1) \delta _H k_L+(\gamma +1) k_H \delta _L+(\kappa -1) k_H k_L)]
\end{align}
We note that $c_5 = \mathcal{O}(\delta_L, \delta_H)$ so that equation \eqref{eqn: L quintic appendix} is a singular perturbation problem in the limit $\delta_L \sim \delta_H \ll 1$. 

\section{Deriving closed-form solutions for $m_j^\star$}

We derive the formulae \eqref{eqn: m_j formula} by using the method of generating functions. We set:
\begin{align}
    f(z,t) &:= \sum_{j = 0}^J m_j(t) \, z^j, & 0 \leq z \leq 1. \label{eqn: gen func def}
\end{align}
By differentiating equation \eqref{eqn: gen func def} with respect to $t$ and substituting the ODEs \eqref{eqn: m0 nondim}-\eqref{eqn: mJ nondim}, we find that $f(z,t)$ satisfies the following PDE:
\begin{align}
    \frac{\partial f}{\partial t} +  (1-z) \Big[ k_L L^\star \Big( Jf - z \frac{\partial f}{\partial z} \Big) - k_H H^\star \frac{\partial f}{\partial z} \Big] = 1- (1+\gamma) f. \label{eqn: f pde}
\end{align}
Equation \eqref{eqn: f pde} is subject to the boundary condition:
\begin{align}
    f(1,t) = M(t). \label{eqn: f bcond}
\end{align}
At steady state equations \eqref{eqn: f pde} and \eqref{eqn: f bcond} reduce to the following non-autonomous ODE initial value problem:
\begin{align}
    &(1-z) \Big[ k_L L^\star \Big( Jf - z \frac{d f}{d z} \Big) - k_H H^\star \frac{d f}{d z} \Big] = 1- (1+\gamma) f. \\
 &f(1) = M,
\end{align}
which can be solved in terms of the hypergeometric function ${}_2F_1$:
\begin{align}
    f(z) = M^\star \, {}_2F_1 \Big( 1, -J, p+2, a_c(1-z) \Big), \label{eqn: fsol}
\end{align}
where $a_c$ and $p$ are the following parameter groupings:
\begin{align}
    a_c &:= \frac{k_L L^\star}{k_L L^\star + k_H H^\star}, & p &:= \frac{1+\gamma}{k_L L^\star + k_H H^\star} - 1.  
\end{align}
Series expansion of the solution \eqref{eqn: fsol} about $z = 0$ gives:
\begin{align}
    f(z) &= M^\star \, \sum_{j=0}^J \Bigg( \prod_{\ell=1}^j \frac{a_c (J- \ell + 1)}{(p+1+\ell)} \Bigg) \, {}_2F_1 ( j+1, j-J, 2+j+p, a_c ) \, z^j,
\end{align}
where the summation is finite because each contribution for $j > J$ is zero. Upon comparison with definition \eqref{eqn: gen func def}, we obtain:
\begin{align}
\begin{split}
    m_j^\star &= M^\star \Bigg( \prod_{\ell=1}^j \frac{a_c (J- \ell + 1)}{(p+1+\ell)} \Bigg) \, {}_2F_1 ( j+1, j-J, 2+j+p, a_c ), 
\end{split} \nonumber \\
&= M^\star \Bigg( \prod_{\ell=1}^j \frac{a_c (J- \ell + 1)}{(p+1+\ell)} \Bigg) \sum_{n = 0}^{J-j} \frac{(j+1)^{(n)} (j-J)^{(n)}}{(j+2+p)^{(n)}} \frac{a_c^n}{n!}, \quad 0 \leq j \leq J, \label{eqn: m_j formula}
\end{align}
where $x^{(n)} = x (x+1) \cdots (x+n-1)$ is the $n$-th rising factorial of $x$.

\section{Deviation between discrete and continuum models}

\begin{figure}[h]
    \centering
    \includegraphics[width=0.9\textwidth]{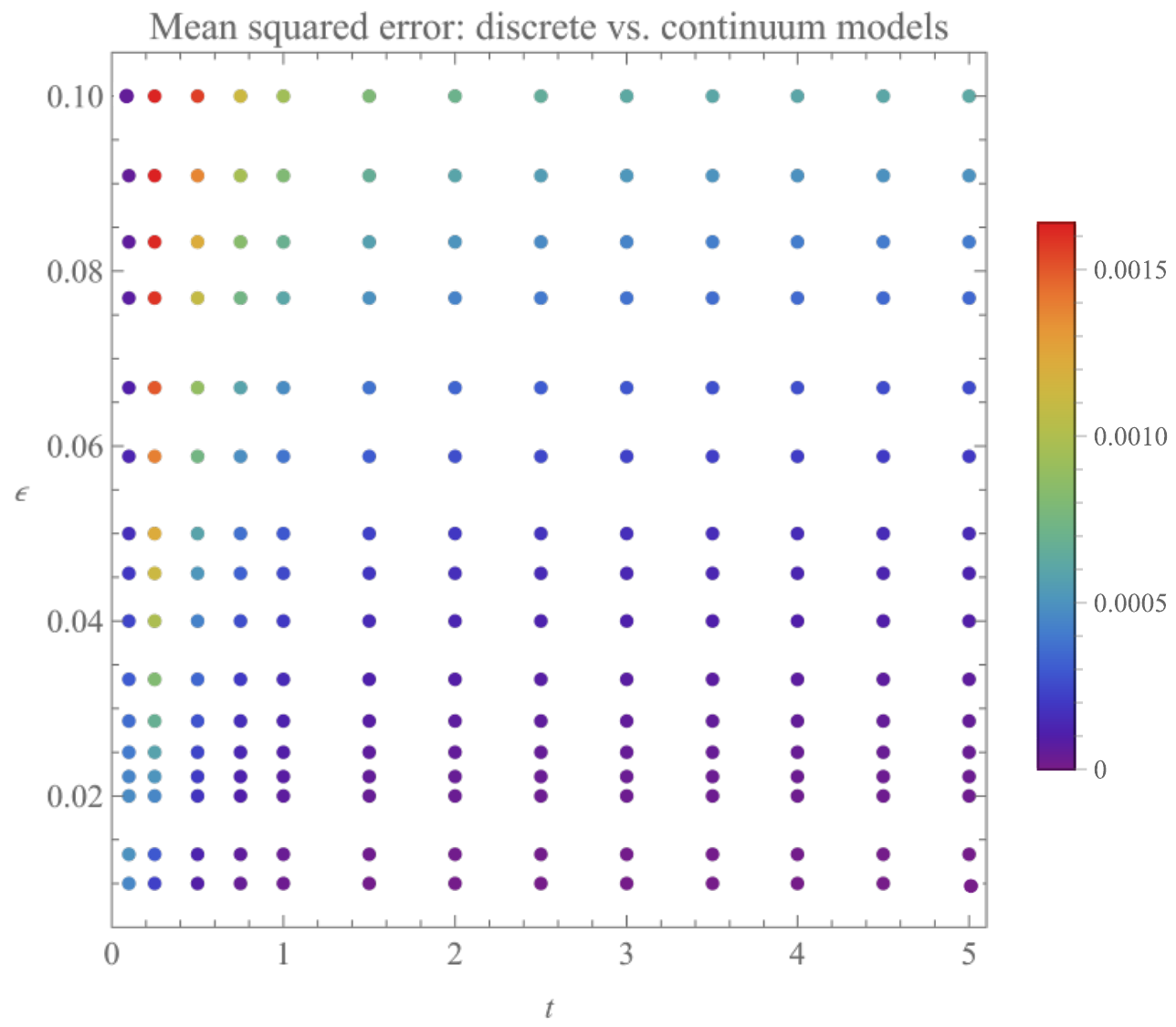}
    \caption{\textbf{Deviation between the discrete and continuum models.} We use equation \eqref{eqn: MSE} to calculate the mean squared error (MSE) across a number of sample times $t$ and values of $\epsilon$. The value of the MSE is indicated by the colour of the corresponding sample point. We use: $\sigma_L = 3$, $k_L = 1$, $k_H H = 0.3$, $\kappa = 30$, $\gamma = 0.25$ and $L_c = 1$.}
    \label{fig:MSE_plot}
\end{figure}

We quantify the deviation between the discrete and continuum models using the mean squared error (MSE), defined in equation \eqref{eqn: MSE}. In Figure \ref{fig:MSE_plot}, we plot the MSE for various values of $\epsilon$ and times $t$. 

\newpage

\bibliographystyle{elsarticle-num} 
\bibliography{cas-refs}

\begin{thebibliography}{10}
\expandafter\ifx\csname url\endcsname\relax
  \def\url#1{\texttt{#1}}\fi
\expandafter\ifx\csname urlprefix\endcsname\relax\def\urlprefix{URL }\fi
\expandafter\ifx\csname href\endcsname\relax
  \def\href#1#2{#2} \def\path#1{#1}\fi

\bibitem{back2019inflammation}
M.~B{\"a}ck, A.~Yurdagul, I.~Tabas, K.~{\"O}{\"o}rni, P.~T. Kovanen,
  Inflammation and its resolution in atherosclerosis: mediators and therapeutic
  opportunities, Nature Reviews Cardiology 16~(7) (2019) 389--406.
\newblock \href {https://doi.org/10.1038/s41569-019-0169-2}
  {\path{doi:10.1038/s41569-019-0169-2}}.

\bibitem{tabas2016macrophage}
I.~Tabas, K.~E. Bornfeldt, Macrophage phenotype and function in different
  stages of atherosclerosis, Circulation research 118~(4) (2016) 653--667.
\newblock \href {https://doi.org/10.1161/CIRCRESAHA.115.306256}
  {\path{doi:10.1161/CIRCRESAHA.115.306256}}.

\bibitem{boren2020low}
J.~Boren, M.~J. Chapman, R.~M. Krauss, C.~J. Packard, J.~F. Bentzon, C.~J.
  Binder, M.~J. Daemen, L.~L. Demer, R.~A. Hegele, S.~J. Nicholls, et~al.,
  Low-density lipoproteins cause atherosclerotic cardiovascular disease:
  pathophysiological, genetic, and therapeutic insights: a consensus statement
  from the european atherosclerosis society consensus panel, European heart
  journal 41~(24) (2020) 2313--2330.
\newblock \href {https://doi.org/10.1093/eurheartj/ehz962}
  {\path{doi:10.1093/eurheartj/ehz962}}.

\bibitem{razeghian2022immune}
I.~Razeghian-Jahromi, A.~Karimi~Akhormeh, M.~Razmkhah, M.~J. Zibaeenezhad,
  Immune system and atherosclerosis: Hostile or friendly relationship,
  International Journal of Immunopathology and Pharmacology 36 (2022)
  03946320221092188.
\newblock \href {https://doi.org/10.1177/03946320221092188}
  {\path{doi:10.1177/03946320221092188}}.

\bibitem{guyton1996development}
J.~R. Guyton, K.~F. Klemp, Development of the lipid-rich core in human
  atherosclerosis, Arteriosclerosis, thrombosis, and vascular biology 16~(1)
  (1996) 4--11.
\newblock \href {https://doi.org/10.1161/01.ATV.16.1.4}
  {\path{doi:10.1161/01.ATV.16.1.4}}.

\bibitem{alique2015ldl}
M.~Alique, C.~Luna, J.~Carracedo, R.~Ram{\'\i}rez, Ldl biochemical
  modifications: a link between atherosclerosis and aging, Food \& nutrition
  research 59~(1) (2015) 29240.
\newblock \href {https://doi.org/10.3402/fnr.v59.29240}
  {\path{doi:10.3402/fnr.v59.29240}}.

\bibitem{zent2017maxed}
C.~S. Zent, M.~R. Elliott, Maxed out macs: physiologic cell clearance as a
  function of macrophage phagocytic capacity, The FEBS journal 284~(7) (2017)
  1021--1039.
\newblock \href {https://doi.org/10.1111/febs.13961}
  {\path{doi:10.1111/febs.13961}}.

\bibitem{remmerie2018macrophages}
A.~Remmerie, C.~L. Scott, Macrophages and lipid metabolism, Cellular immunology
  330 (2018) 27--42.
\newblock \href {https://doi.org/10.1016/j.cellimm.2018.01.020}
  {\path{doi:10.1016/j.cellimm.2018.01.020}}.

\bibitem{pinney2020macrophage}
J.~J. Pinney, F.~Rivera-Escalera, C.~C. Chu, H.~E. Whitehead, K.~R. VanDerMeid,
  A.~M. Nelson, M.~C. Barbeau, C.~S. Zent, M.~R. Elliott, Macrophage hypophagia
  as a mechanism of innate immune exhaustion in mab-induced cell clearance,
  Blood 136~(18) (2020) 2065--2079.
\newblock \href {https://doi.org/10.1182/blood.2020005571}
  {\path{doi:10.1182/blood.2020005571}}.

\bibitem{kojima2017role}
Y.~Kojima, I.~L. Weissman, N.~J. Leeper, The role of efferocytosis in
  atherosclerosis, Circulation 135~(5) (2017) 476--489.
\newblock \href {https://doi.org/10.1161/CIRCULATIONAHA.116.025684}
  {\path{doi:10.1161/CIRCULATIONAHA.116.025684}}.

\bibitem{yurdagul2018mechanisms}
A.~Yurdagul~Jr, A.~C. Doran, B.~Cai, G.~Fredman, I.~A. Tabas, Mechanisms and
  consequences of defective efferocytosis in atherosclerosis, Frontiers in
  cardiovascular medicine 4 (2018) 86.
\newblock \href {https://doi.org/10.3389/fcvm.2017.00086}
  {\path{doi:10.3389/fcvm.2017.00086}}.

\bibitem{avgerinos2019mathematical}
N.~A. Avgerinos, P.~Neofytou, Mathematical modelling and simulation of
  atherosclerosis formation and progress: a review, Annals of Biomedical
  Engineering 47~(8) (2019) 1764--1785.
\newblock \href {https://doi.org/10.1007/s10439-019-02268-3}
  {\path{doi:10.1007/s10439-019-02268-3}}.

\bibitem{el2019mathematical}
N.~El~Khatib, O.~Kafi, A.~Sequeira, S.~Simakov, Y.~Vassilevski, V.~Volpert,
  Mathematical modelling of atherosclerosis, Mathematical modelling of natural
  phenomena 14~(6) (2019) 603.
\newblock \href {https://doi.org/10.1051/mmnp/2019050}
  {\path{doi:10.1051/mmnp/2019050}}.

\bibitem{calvez2009mathematical}
V.~Calvez, A.~Ebde, N.~Meunier, A.~Raoult, Mathematical modelling of the
  atherosclerotic plaque formation, in: Esaim: Proceedings, Vol.~28, EDP
  Sciences, 2009, pp. 1--12.
\newblock \href {https://doi.org/10.1051/proc/2009036}
  {\path{doi:10.1051/proc/2009036}}.

\bibitem{chalmers2017nonlinear}
A.~D. Chalmers, C.~A. Bursill, M.~R. Myerscough, Nonlinear dynamics of early
  atherosclerotic plaque formation may determine the efficacy of high density
  lipoproteins (hdl) in plaque regression, PLoS One 12~(11) (2017) e0187674.
\newblock \href {https://doi.org/10.1371/journal.pone.0187674}
  {\path{doi:10.1371/journal.pone.0187674}}.

\bibitem{silva2020modeling}
T.~Silva, W.~J{\"a}ger, M.~Neuss-Radu, A.~Sequeira, Modeling of the early stage
  of atherosclerosis with emphasis on the regulation of the endothelial
  permeability, Journal of Theoretical Biology 496 (2020) 110229.
\newblock \href {https://doi.org/10.1016/j.jtbi.2020.110229}
  {\path{doi:10.1016/j.jtbi.2020.110229}}.

\bibitem{ford2019lipid}
H.~Z. Ford, H.~M. Byrne, M.~R. Myerscough, A lipid-structured model for
  macrophage populations in atherosclerotic plaques, Journal of Theoretical
  Biology 479 (2019) 48--63.
\newblock \href {https://doi.org/10.1016/j.jtbi.2019.07.003}
  {\path{doi:10.1016/j.jtbi.2019.07.003}}.

\bibitem{chambers2022lipid}
K.~L. Chambers, M.~G. Watson, M.~R. Myerscough, A lipid-structured mathematical
  model of atherosclerosis with macrophage proliferation, arXiv preprint
  arXiv:2205.04715 (2022).
\newblock \href {https://doi.org/10.48550/arXiv.2205.04715}
  {\path{doi:10.48550/arXiv.2205.04715}}.

\bibitem{watson2022lipid}
M.~G. Watson, K.~L. Chambers, M.~R. Myerscough, A lipid-structured model of
  atherosclerotic plaque macrophages with lipid-dependent kinetics, arXiv
  preprint arXiv:2205.05285 (2022).
\newblock \href {https://doi.org/10.48550/arXiv.2205.05285}
  {\path{doi:10.48550/arXiv.2205.05285}}.

\bibitem{meunier2019mathematical}
N.~Meunier, N.~Muller, Mathematical study of an inflammatory model for
  atherosclerosis: a nonlinear renewal equation, Acta Applicandae Mathematicae
  161~(1) (2019) 107--126.
\newblock \href {https://doi.org/10.1007/s10440-018-0206-x}
  {\path{doi:10.1007/s10440-018-0206-x}}.

\bibitem{williams2020limited}
J.~W. Williams, K.~Zaitsev, K.-W. Kim, S.~Ivanov, B.~T. Saunders, P.~R.
  Schrank, K.~Kim, A.~Elvington, S.~H. Kim, C.~G. Tucker, et~al., Limited
  proliferation capacity of aortic intima resident macrophages requires
  monocyte recruitment for atherosclerotic plaque progression, Nature
  immunology 21~(10) (2020) 1194--1204.
\newblock \href {https://doi.org/10.1038/s41590-020-0768-4}
  {\path{doi:10.1038/s41590-020-0768-4}}.

\bibitem{ford2019efferocytosis}
H.~Z. Ford, L.~Zeboudj, G.~S. Purvis, A.~Ten~Bokum, A.~E. Zarebski, J.~A. Bull,
  H.~M. Byrne, M.~R. Myerscough, D.~R. Greaves, Efferocytosis perpetuates
  substance accumulation inside macrophage populations, Proceedings of the
  Royal Society B 286~(1904) (2019) 20190730.
\newblock \href {https://doi.org/10.1098/rspb.2019.0730}
  {\path{doi:10.1098/rspb.2019.0730}}.

\bibitem{taefehshokr2021rab}
N.~Taefehshokr, C.~Yin, B.~Heit, Rab gtpases in the differential processing of
  phagocytosed pathogens versus efferocytosed apoptotic cells, Histol.
  Histopathol 36 (2021) 123--135.
\newblock \href {https://doi.org/10.14670/HH-18-252}
  {\path{doi:10.14670/HH-18-252}}.

\bibitem{westman2020phagocytosis}
J.~Westman, S.~Grinstein, P.~E. Marques, Phagocytosis of necrotic debris at
  sites of injury and inflammation, Frontiers in immunology 10 (2020) 3030.
\newblock \href {https://doi.org/10.3389/fimmu.2019.03030}
  {\path{doi:10.3389/fimmu.2019.03030}}.

\bibitem{thon2018quantitative}
M.~P. Thon, H.~Z. Ford, M.~W. Gee, M.~R. Myerscough, A quantitative model of
  early atherosclerotic plaques parameterized using in vitro experiments,
  Bulletin of mathematical biology 80~(1) (2018) 175--214.
\newblock \href {https://doi.org/10.1007/s11538-017-0367-1}
  {\path{doi:10.1007/s11538-017-0367-1}}.

\bibitem{angelovich2017quantification}
T.~A. Angelovich, A.~C. Hearps, A.~Maisa, T.~Kelesidis, A.~Jaworowski,
  Quantification of monocyte transmigration and foam cell formation from
  individuals with chronic inflammatory conditions, JoVE (Journal of Visualized
  Experiments) (2017) e56293\href {https://doi.org/10.3791/56293}
  {\path{doi:10.3791/56293}}.

\bibitem{ghattas2013monocytes}
A.~Ghattas, H.~R. Griffiths, A.~Devitt, G.~Y. Lip, E.~Shantsila, Monocytes in
  coronary artery disease and atherosclerosis: where are we now?, Journal of
  the American College of Cardiology 62~(17) (2013) 1541--1551.
\newblock \href {https://doi.org/10.1016/j.jacc.2013.07.043}
  {\path{doi:10.1016/j.jacc.2013.07.043}}.

\bibitem{bruckert2011lowering}
E.~Bruckert, D.~Rosenbaum, Lowering ldl-cholesterol through diet: potential
  role in the statin era, Current opinion in lipidology 22~(1) (2011) 43--48.
\newblock \href {https://doi.org/10.1097/MOL.0b013e328340b8e7}
  {\path{doi:10.1097/MOL.0b013e328340b8e7}}.

\bibitem{1F1}
Kummer confluent hypergeometric function,
  \url{https://functions.wolfram.com/HypergeometricFunctions/Hypergeometric1F1/}.

\bibitem{HermiteH}
Hermite hypergeometric function,
  \url{https://functions.wolfram.com/HypergeometricFunctions/HermiteHGeneral/}.

\bibitem{ge2022efferocytosis}
Y.~Ge, M.~Huang, Y.-m. Yao, Efferocytosis and its role in inflammatory
  disorders, Frontiers in Cell and Developmental Biology 10 (2022).
\newblock \href {https://doi.org/10.3389/fcell.2022.839248}
  {\path{doi:10.3389/fcell.2022.839248}}.

\bibitem{yin2021cellular}
C.~Yin, B.~Heit, Cellular responses to the efferocytosis of apoptotic cells,
  Frontiers in Immunology 12 (2021) 631714.
\newblock \href {https://doi.org/10.3389/fimmu.2021.631714}
  {\path{doi:10.3389/fimmu.2021.631714}}.

\bibitem{chen2019lysophosphatidic}
L.~Chen, J.~Zhang, X.~Yang, Y.~Liu, X.~Deng, C.~Yu, Lysophosphatidic acid
  decreased macrophage foam cell migration correlated with downregulation of
  fucosyltransferase 8 via hnf1$\alpha$, Atherosclerosis 290 (2019) 19--30.
\newblock \href {https://doi.org/10.1016/j.atherosclerosis.2019.09.001}
  {\path{doi:10.1016/j.atherosclerosis.2019.09.001}}.

\bibitem{robbins2013local}
C.~S. Robbins, I.~Hilgendorf, G.~F. Weber, I.~Theurl, Y.~Iwamoto, J.-L.
  Figueiredo, R.~Gorbatov, G.~K. Sukhova, L.~Gerhardt, D.~Smyth, et~al., Local
  proliferation dominates lesional macrophage accumulation in atherosclerosis,
  Nature medicine 19~(9) (2013) 1166--1172.
\newblock \href {https://doi.org/10.1038/nm.3258} {\path{doi:10.1038/nm.3258}}.

\bibitem{celora2023spatio}
G.~L. Celora, H.~M. Byrne, P.~Kevrekidis, Spatio-temporal modelling of
  phenotypic heterogeneity in tumour tissues and its impact on radiotherapy
  treatment, Journal of Theoretical Biology 556 (2023) 111248.
\newblock \href {https://doi.org/10.1016/j.jtbi.2022.111248}
  {\path{doi:10.1016/j.jtbi.2022.111248}}.

\bibitem{fiandaca2022phenotype}
G.~Fiandaca, S.~Bernardi, M.~Scianna, M.~E. Delitala, A phenotype-structured
  model to reproduce the avascular growth of a tumor and its interaction with
  the surrounding environment, Journal of Theoretical Biology 535 (2022)
  110980.
\newblock \href {https://doi.org/10.1016/j.jtbi.2021.110980}
  {\path{doi:10.1016/j.jtbi.2021.110980}}.

\bibitem{kim2018transcriptome}
K.~Kim, D.~Shim, J.~S. Lee, K.~Zaitsev, J.~W. Williams, K.-W. Kim, M.-Y. Jang,
  H.~Seok~Jang, T.~J. Yun, S.~H. Lee, et~al., Transcriptome analysis reveals
  nonfoamy rather than foamy plaque macrophages are proinflammatory in
  atherosclerotic murine models, Circulation research 123~(10) (2018)
  1127--1142.
\newblock \href {https://doi.org/10.1161/CIRCRESAHA.118.312804}
  {\path{doi:10.1161/CIRCRESAHA.118.312804}}.

\end{thebibliography}





% \end{thebibliography}
\end{document}